\documentclass[trackchanges]{aastex701}

\usepackage{graphicx}   
\usepackage{subcaption} 
\usepackage{booktabs} 
\usepackage{xcolor}

\begin{document}

\title{M-EPDet: Real-Time Real-Bogus Classification and Transient Candidate Judgement for the EP-WXT Pipeline via Multi-Modal Data}

\author[sname=Chen,gname=Lang]{Lang Chen}
\affiliation{National Astronomical Observatories, Chinese Academy of Sciences, Beijing 100101, China}
\affiliation{University of Chinese Academy of Sciences, Beijing 100049, China}
\affiliation{National Astronomical Data Center, Beijing 100101, China}
\email{chenlang@bao.ac.cn}  

\author[sname=Xu,gname=Yunfei]{Yunfei Xu}
\affiliation{National Astronomical Observatories, Chinese Academy of Sciences, Beijing 100101, China}
\affiliation{University of Chinese Academy of Sciences, Beijing 100049, China}
\affiliation{National Astronomical Data Center, Beijing 100101, China}
\email[show]{xuyf@nao.cas.cn}  

\author[sname=Zhang,gname=Zhen]{Zhen Zhang}
\affiliation{National Astronomical Observatories, Chinese Academy of Sciences, Beijing 100101, China}
\affiliation{University of Chinese Academy of Sciences, Beijing 100049, China}
\affiliation{National Astronomical Data Center, Beijing 100101, China}
\email{zhangzhen@nao.cas.cn}  

\author[sname=Li,gname=Dongyue]{Dongyue Li}
\affiliation{National Astronomical Observatories, Chinese Academy of Sciences, Beijing 100101, China}
\email{dyli@nao.cas.cn}  

\author[sname=Sun,gname=Hui]{Hui Sun}
\affiliation{National Astronomical Observatories, Chinese Academy of Sciences, Beijing 100101, China}
\email{hsun@nao.cas.cn}  

\author[sname=Liu,gname=Yuan]{Yuan Liu}
\affiliation{National Astronomical Observatories, Chinese Academy of Sciences, Beijing 100101, China}
\email{liuyuan@bao.ac.cn}  

\author[sname=Cui,gname=Chenzhou]{Chenzhou Cui}
\affiliation{National Astronomical Observatories, Chinese Academy of Sciences, Beijing 100101, China}
\affiliation{University of Chinese Academy of Sciences, Beijing 100049, China}
\affiliation{National Astronomical Data Center, Beijing 100101, China}
\email[show]{ccz@bao.ac.cn} 

\author[sname=Xie,gname=Jinhui]{Jinhui Xie}
\affiliation{National Astronomical Observatories, Chinese Academy of Sciences, Beijing 100101, China}
\affiliation{University of Chinese Academy of Sciences, Beijing 100049, China}
\affiliation{National Astronomical Data Center, Beijing 100101, China}
\email{xiejinhui22@mails.ucas.ac.cn}  

\author[sname=Zuo,gname=XiaoXiong]{XiaoXiong Zuo}
\affiliation{National Astronomical Observatories, Chinese Academy of Sciences, Beijing 100101, China}
\affiliation{University of Chinese Academy of Sciences, Beijing 100049, China}
\affiliation{National Astronomical Data Center, Beijing 100101, China}
\email{zuoxx@bao.ac.cn}  

\author[sname=Wei,gname=Shirui]{Shirui Wei}
\affiliation{National Astronomical Observatories, Chinese Academy of Sciences, Beijing 100101, China}
\affiliation{University of Chinese Academy of Sciences, Beijing 100049, China}
\affiliation{National Astronomical Data Center, Beijing 100101, China}
\email{weisr@bao.ac.cn}   

\author[sname=Shao,gname=Wujun]{Wujun Shao}
\affiliation{National Astronomical Observatories, Chinese Academy of Sciences, Beijing 100101, China}
\affiliation{University of Chinese Academy of Sciences, Beijing 100049, China}
\affiliation{National Astronomical Data Center, Beijing 100101, China}
\email{shaowj@bao.ac.cn}   

\correspondingauthor{Yunfei Xu, Chenzhou Cui}

\begin{abstract}

The Wide-field X-ray Telescope (WXT) onboard the Einstein Probe (EP) produces a large post-detection candidate stream in which genuine astrophysical sources coexist with instrumental artifacts and Cosmic Ray events. We present M-EPDet, a three-step post-detection framework for real-time candidate vetting in EP-WXT lobster-eye Micro-pore Optics (MPO) data. The framework combines a ResNet-based Arm filter, a dual-branch temporal-spectral Cosmic Ray filter, and a background-aware Bayesian Blocks module for single-exposure variability screening.

Using on-orbit EP-WXT observations, we report decoupled metrics for the cascading system. M-EPDet achieves a Real-Bogus Recall of 98.31\% ($98.53\% \times 99.78\%$) for genuine astrophysical sources, together with rejection rates of 92.99\% for instrumental artifacts and 98.18\% for Cosmic Ray events. In the final step, the Bayesian Blocks module flags 0.75\% of the post-filtration observations, corresponding to a 99.25\% reduction in candidate volume. The system is deployed in the EP-WXT pipeline as a lightweight real-time service, reducing the manual-inspection burden in candidate vetting.

\end{abstract}



\section{Introduction}
\label{sec:intro}

X-ray time-domain astronomy provides an important observational window on a wide range of high-energy and transient phenomena. Unlike the relatively static optical or radio skies, the X-ray sky is highly dynamic, teeming with transients and variable sources. According to \cite{EP_White_Paper}, the soft X-ray band (0.5--4 keV) is critical for detecting two key categories of astrophysical events: flares from tidal disruption events (TDEs) caused by dormant black holes, and electromagnetic counterparts to gravitational wave events such as binary neutron star mergers. Furthermore, capturing high-redshift Gamma-Ray Bursts (GRBs) and supernova shock breakouts is essential for probing fundamental physics, including early universe star formation and the origin of heavy elements \cite{EP_White_Paper}. To capture these faint and short-lived phenomena, next-generation X-ray telescopes generally require both wide field of view and high sensitivity. The Einstein Probe (EP) is a time-domain observatory designed for this purpose. Its primary payload, the Wide-field X-ray Telescope (WXT), utilizes bio-inspired Micro-pore Optics (MPO) technology to achieve an instantaneous field of view (FoV) of about 3600 square degrees, enabling wide-field and sensitive sky monitoring \cite{EP_Paper}.
However, this optical design also introduces data-processing challenges that differ from those of conventional focusing telescopes. The lobster-eye MPO imaging principle produces a cruciform Point Spread Function (PSF), and in on-orbit observations its extended Arms become a major source of background interference \cite{EP_White_Paper,Ling2023LEIA,Cheng2025WXT,Zhang2022LEIA}. In the EP-WXT pipeline, genuine astrophysical sources therefore coexist with instrumental Arm artifacts and Cosmic Ray events in a large post-detection candidate stream.

Machine-learning-based Real-Bogus filtering has become standard in optical time-domain surveys, and related approaches have also been explored in X-ray applications \cite{SanchezSaez2021,Shah2025,Killestein2021,Goode2022,Tranin2022,Ruiz2024,Dillmann2025,Jia2023}. However, methods developed for approximately Gaussian-like point sources do not transfer directly to EP-WXT lobster-eye data, and approaches based on pathfinder observations or hand-crafted features may not fully capture the complexity of on-orbit backgrounds \cite{Zuo2024}.

To address this problem, we propose M-EPDet, a hierarchical multi-modal post-detection framework for real-time candidate vetting in the EP-WXT pipeline. The framework consists of three sequential components: an Arm Filter based on ResNet in the spatial domain, a dual-branch Cosmic Ray Filter in the temporal-spectral domain, and a Bayesian-Blocks-based Variability Screening module for candidate prioritization. The remainder of this paper is organized as follows. Section~\ref{sec:data} describes the construction of the multi-modal dataset, Section~\ref{sec:method} presents the three-step M-EPDet framework, Section~\ref{sec:results} reports the step-by-step experimental results, Section~\ref{sec:application} discusses deployment and practical limitations, and Section~\ref{sec:conc} summarizes the main conclusions.

\section{Construction of Multi-modal Dataset using EP-WXT Observational Data}
\label{sec:data}

This section details the data structure of the Wide-field X-ray Telescope (WXT) onboard the Einstein Probe (EP) and the construction of the multi-modal dataset used in this study.

\subsection{Instrument and Data Products}
The WXT payload consists of 12 independent modules equipped with 48 large-format CMOS detectors, utilizing bio-inspired lobster-eye Micro-pore Optics (MPO). The observational data are processed by the EP-WXT Data Center into three standard levels. This study primarily utilizes Level-2 and Level-3 data products:

\begin{itemize}
    \item Level-1 (Uncleaned Event Data): Raw Pulse Height Amplitude (PHA) and detector coordinates.
    \item Level-2 (Calibrated and Screened Event Data): The primary input for high-precision analysis. Key steps include Gain/CTI correction, coordinate conversion, and event screening. The resulting \texttt{po\_cl.evt} files provide event times at the WXT temporal resolution of 50 ms.
    \item Level-3 (High-level Scientific Products): Standard products including sky images, source catalogs, light curves, and spectra.
\end{itemize}

\subsection{Construction of Ground Truth and Class Distribution}
To ensure high-fidelity labeling for supervised learning, we leveraged operational logs from EP-WXT data processing pipeline, spanning the first year of the mission's scientific operation (July 10, 2024 -- July 10, 2025). 

The ground truth labels are derived directly from the daily verification records of the EP-WXT Transient Assistants (TA). Rather than relying on subjective visual inspection, these classifications represent definitive physical identifications produced by the official EP scientific workflow. As of July 10, 2025, the tightly curated dataset comprises exactly 211,959 labeled entries. Here, an entry is defined at the candidate-observation level, corresponding to an individual candidate observation generated and verified by the operational EP-WXT pipeline, rather than at the unique-source level. Therefore, multiple observations of the same astrophysical source may appear as separate entries in the full dataset. At the full-dataset level, we deliberately preserved the intrinsic class imbalance of the operational data stream. For the step-specific binary tasks in Step 1 and Step 2, however, balanced benchmarks were constructed as described in Section~\ref{subsec:exp_setup}. The dataset is distributed across the three primary categories targeted by our pipeline:
\begin{itemize}
    \item \textbf{Genuine Astrophysical Sources:} $175,644$ samples.
    \item \textbf{Instrumental Arm Artifacts:} $11,710$ samples.
    \item \textbf{Cosmic Ray events:} $24,605$ samples.
\end{itemize}

\subsection{Multi-modal Data Components}
For each candidate event, we constructed a multi-modal feature triplet $(X_{img}, X_{lc}, X_{spec})$ from the following core data products, together with its ground-truth label $Y$ from the EP-WXT verification records.

\begin{enumerate}
    \item Image Cutout ($X_{img}$): A $100 \times 100$ pixel photon-count image cropped from the Level-3 full-frame sky image. Figures \ref{fig:arm_samples} and \ref{fig:other_samples} illustrate representative samples of the ``Arm'' artifacts and the ``Other'' class.
    
    \begin{figure}[htbp]
        \centering
        \begin{minipage}[b]{0.242\textwidth}
            \centering
            \includegraphics[width=\textwidth]{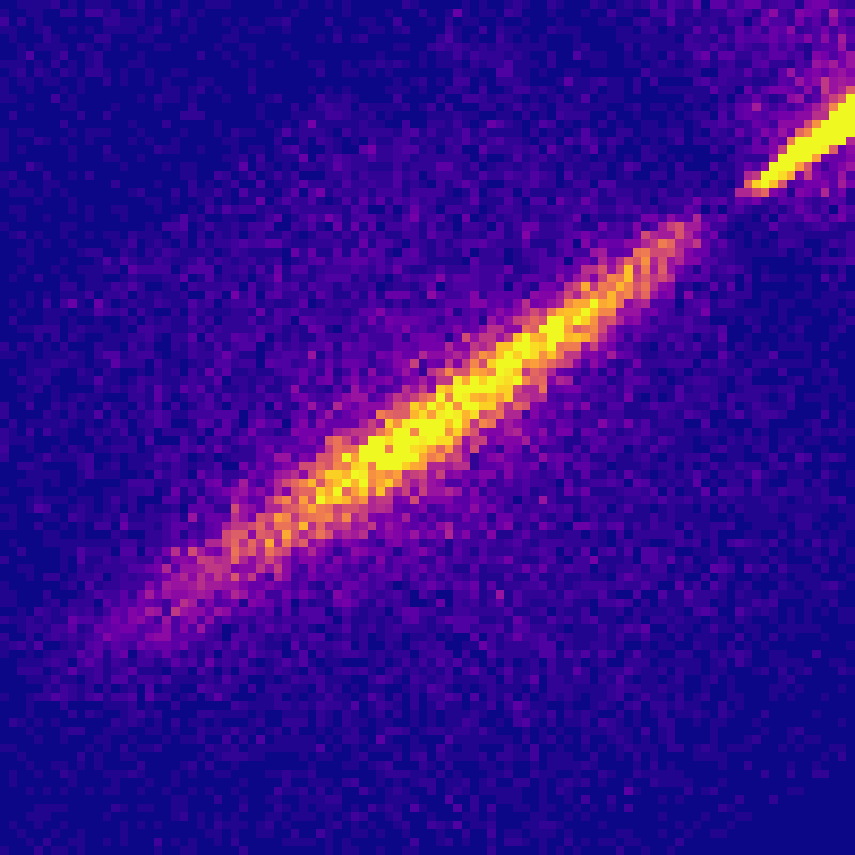}
        \end{minipage}\hfill
        \begin{minipage}[b]{0.242\textwidth}
            \centering
            \includegraphics[width=\textwidth]{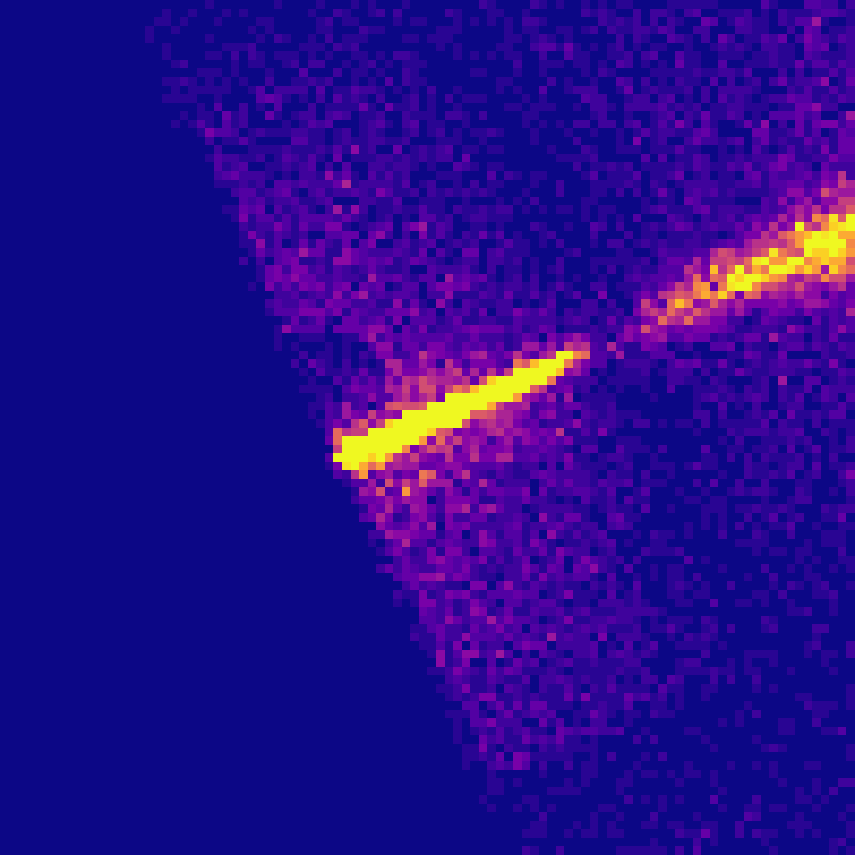}
        \end{minipage}\hfill
        \begin{minipage}[b]{0.242\textwidth}
            \centering
            \includegraphics[width=\textwidth]{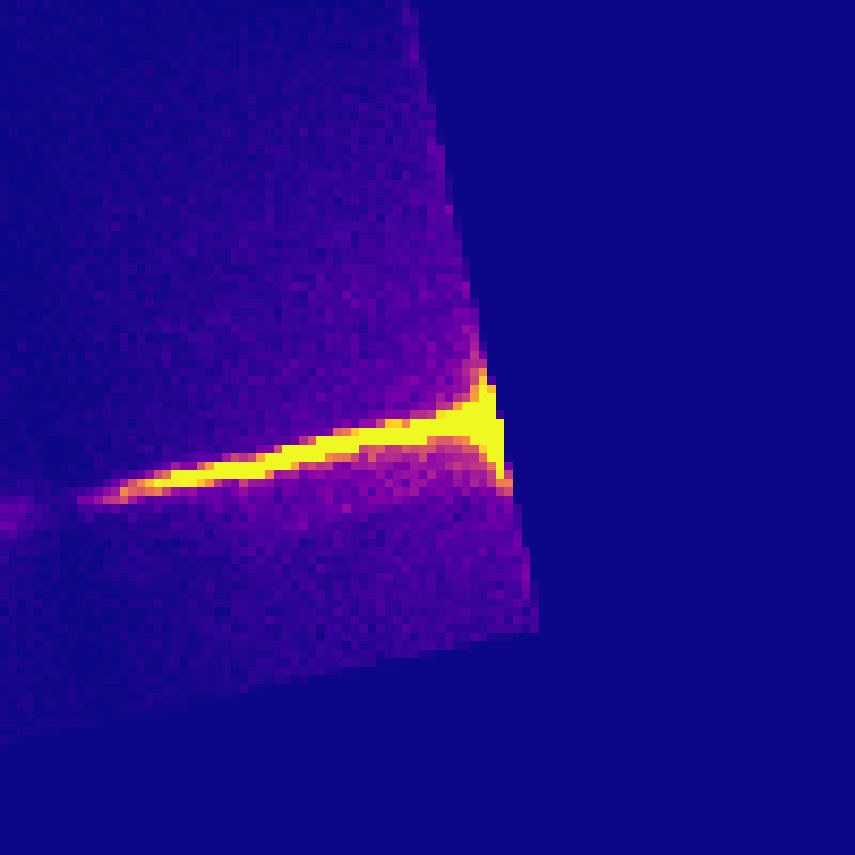}
        \end{minipage}\hfill
        \begin{minipage}[b]{0.242\textwidth}
            \centering
            \includegraphics[width=\textwidth]{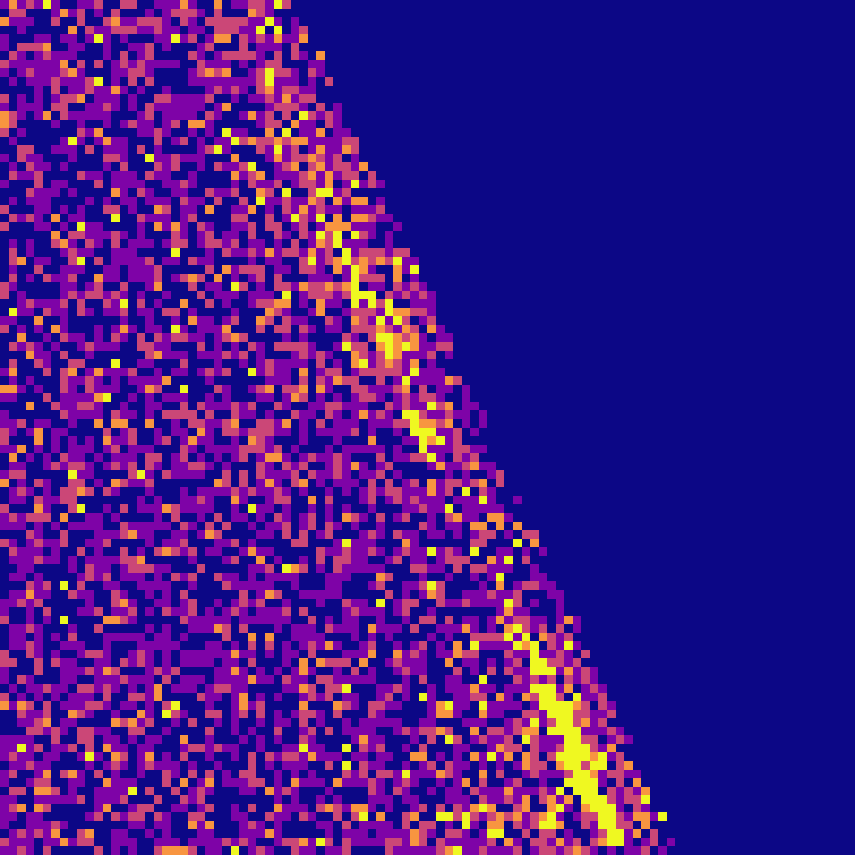}
        \end{minipage}
        \caption{\textbf{Representative samples of instrumental ``Arm'' artifacts.}}
        \label{fig:arm_samples}
    \end{figure}

    \begin{figure}[htbp]
        \centering
        \begin{minipage}[b]{0.242\textwidth}
            \centering
            \includegraphics[width=\textwidth]{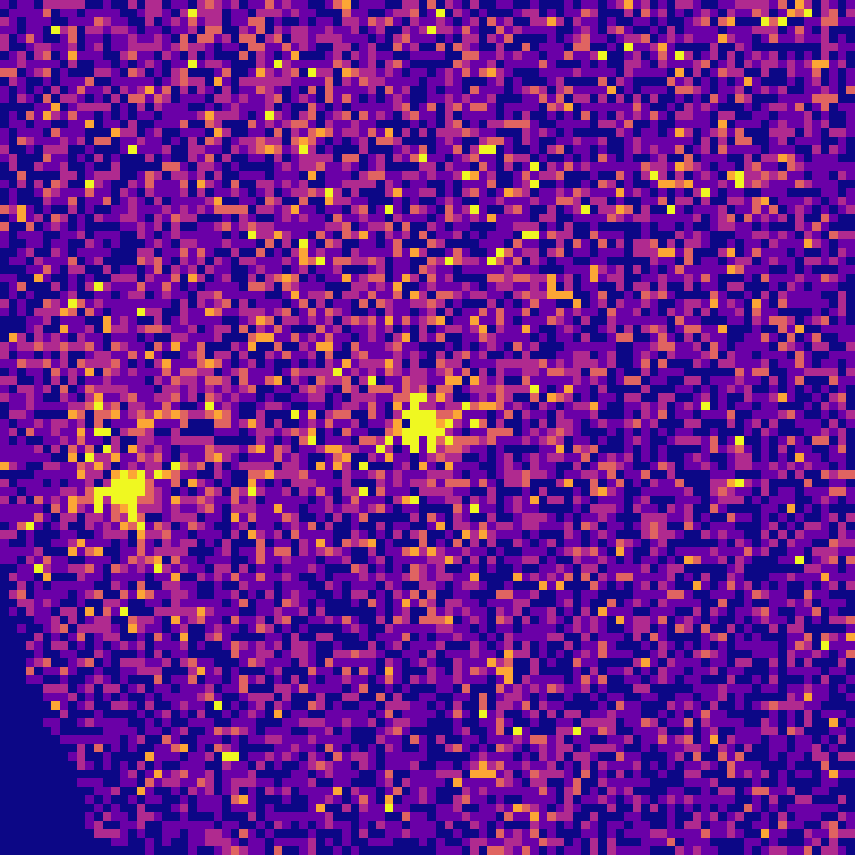}
        \end{minipage}\hfill
        \begin{minipage}[b]{0.242\textwidth}
            \centering
            \includegraphics[width=\textwidth]{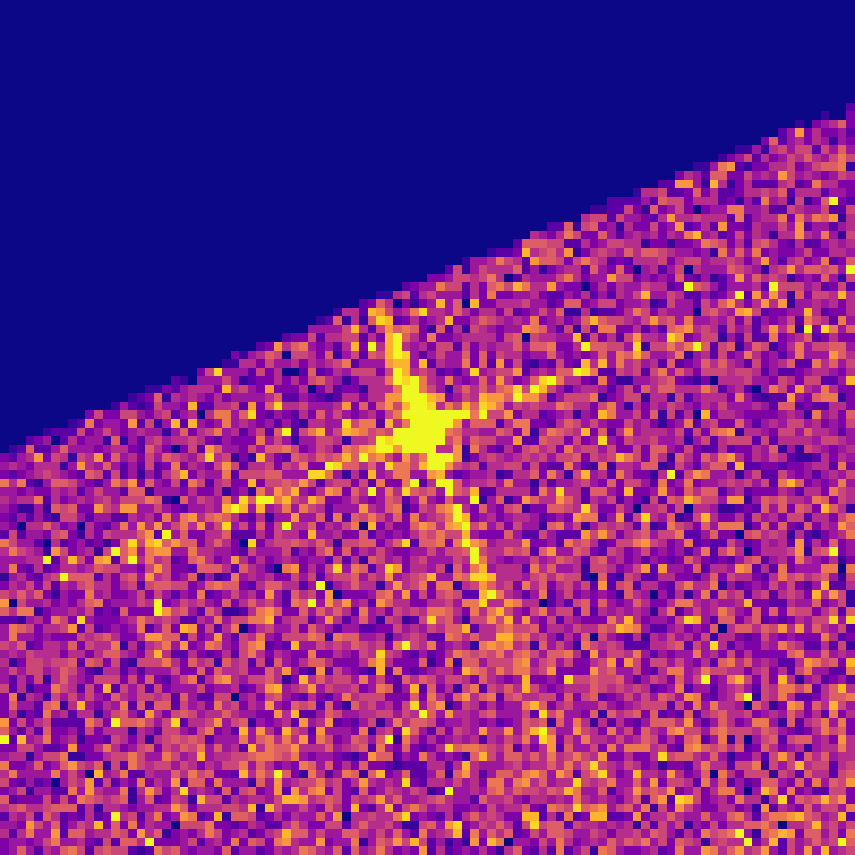}
        \end{minipage}\hfill
        \begin{minipage}[b]{0.242\textwidth}
            \centering
            \includegraphics[width=\textwidth]{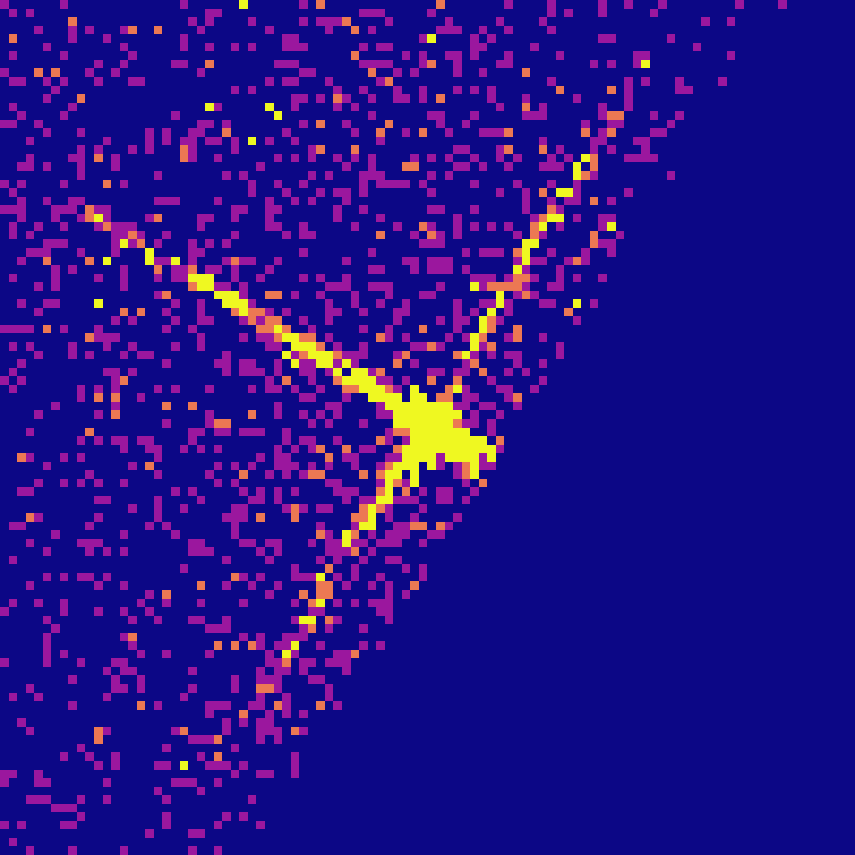}
        \end{minipage}\hfill
        \begin{minipage}[b]{0.242\textwidth}
            \centering
            \includegraphics[width=\textwidth]{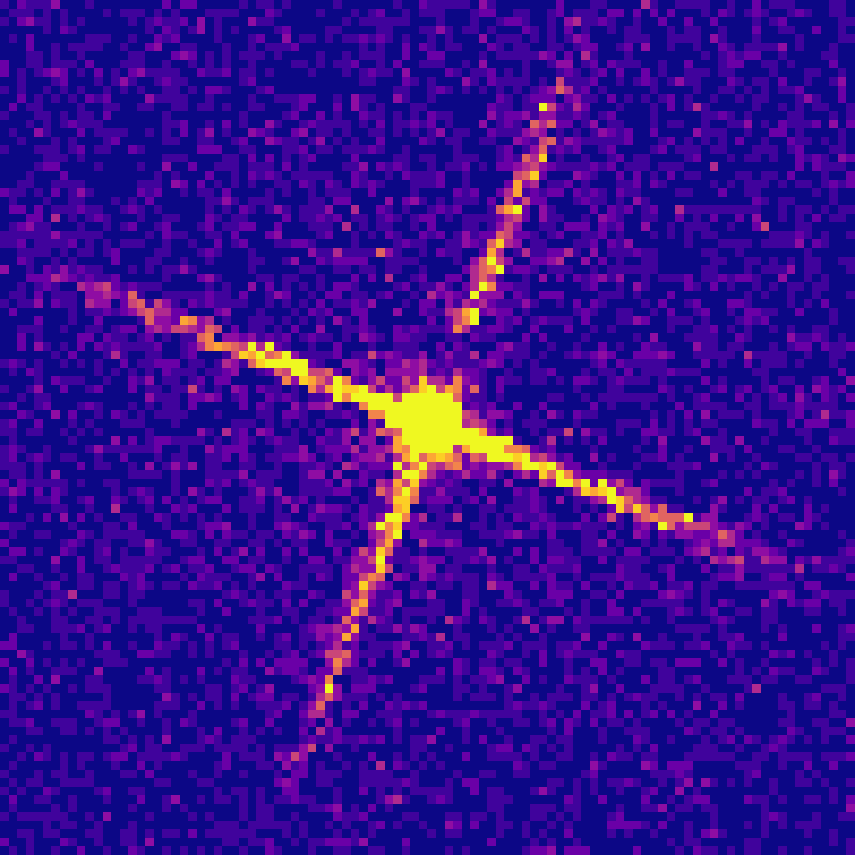}
        \end{minipage}
        \caption{\textbf{Representative samples of the ``Other'' category (Real Sources and Cosmic Ray events).}}
        \label{fig:other_samples}
    \end{figure}

    \item Light Curve ($X_{lc}$) and Energy Spectrum ($X_{spec}$): As shown in Figure \ref{fig:multimodal_comparison}, these two modalities are crucial for distinguishing between Cosmic Ray events and true sources.
    
\begin{figure*}[htbp]
\centering
\gridline{
    \fig{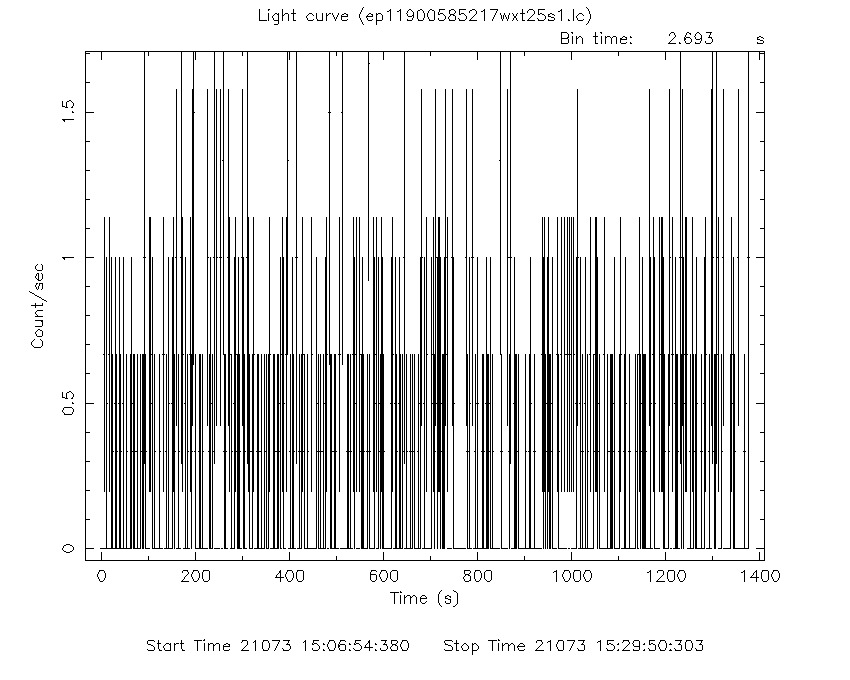}{0.32\textwidth}{(a) Arm: Light Curve}
    \fig{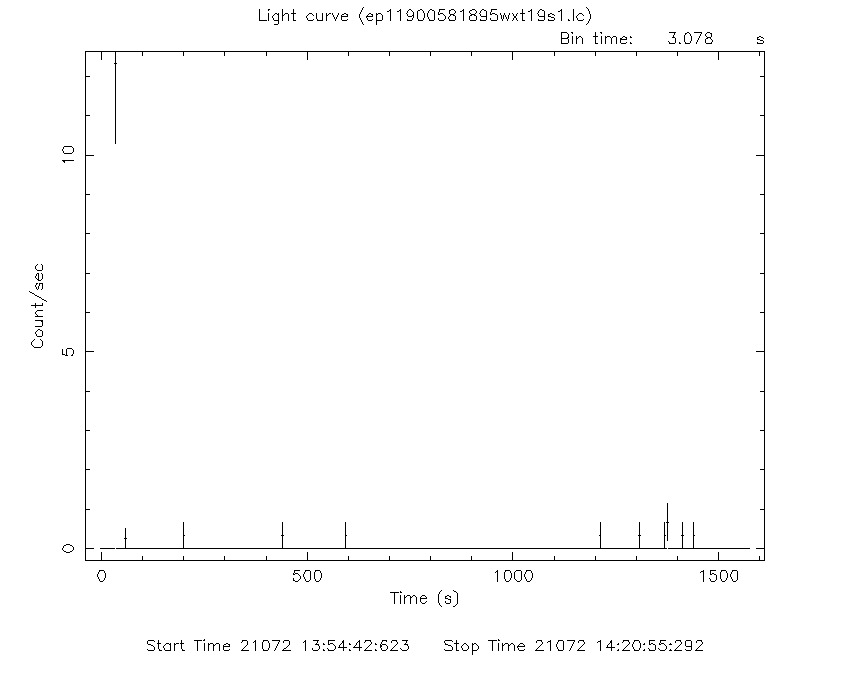}{0.32\textwidth}{(b) Cosmic Ray: Light Curve}
    \fig{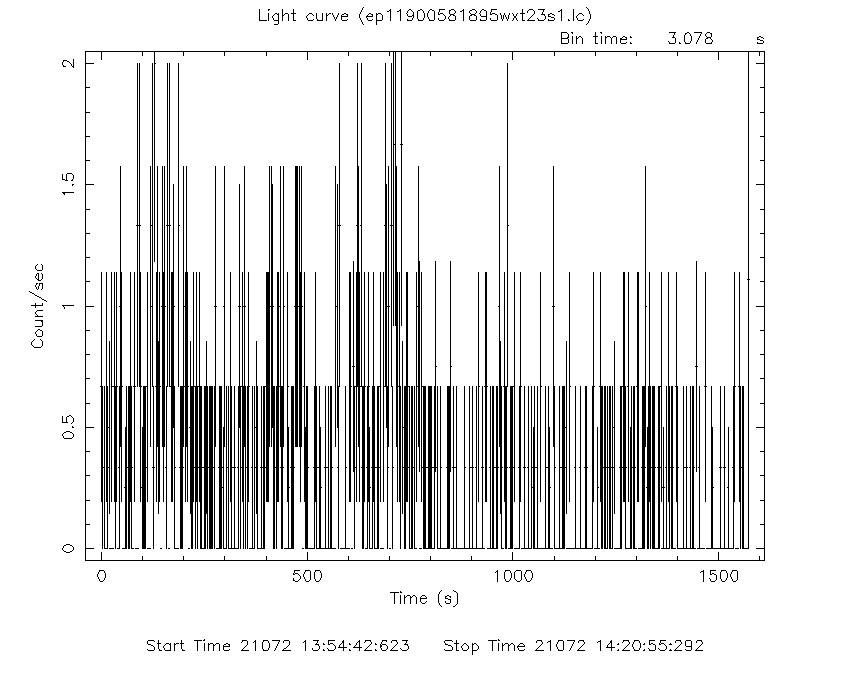}{0.32\textwidth}{(c) Source: Light Curve}
}
\gridline{
    \fig{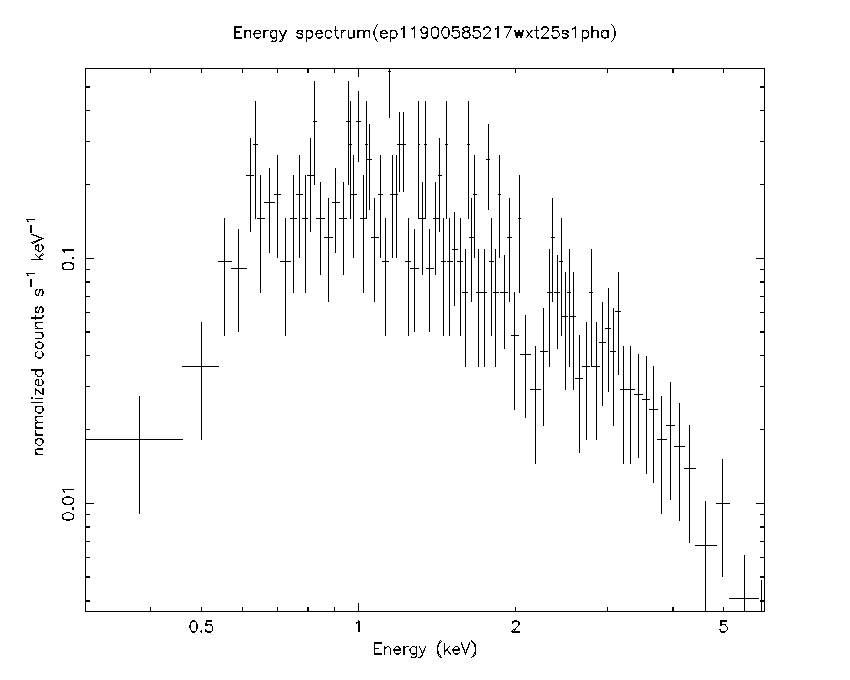}{0.32\textwidth}{(d) Arm: Spectrum}
    \fig{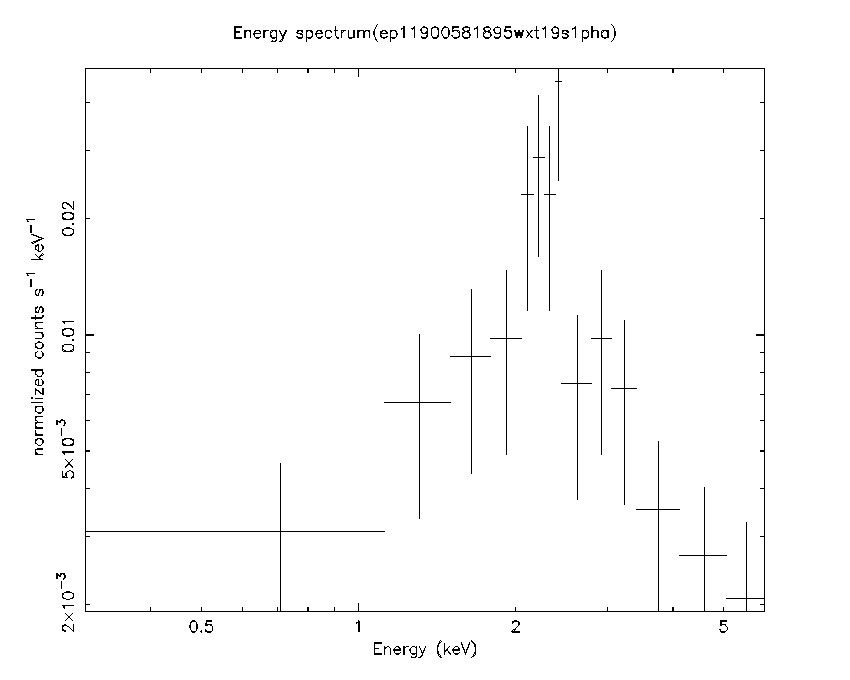}{0.32\textwidth}{(e) Cosmic Ray: Spectrum}
    \fig{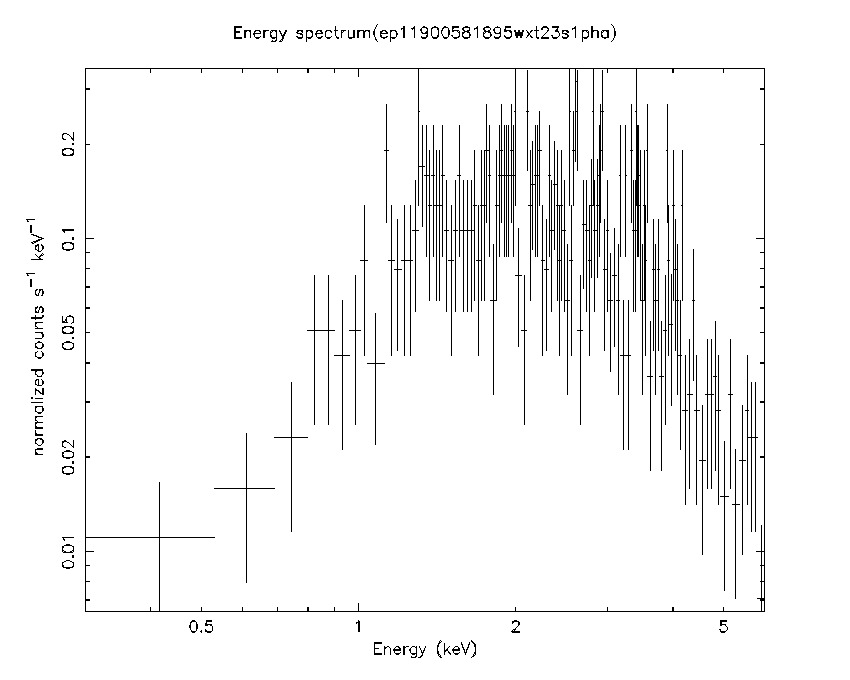}{0.32\textwidth}{(f) Source: Spectrum}
}
\caption{
\textbf{Comparison of temporal and spectral features.}
Top row (Light curves): Arms show noise-like variability, Cosmic Ray events show sharp temporal impulses, and True Sources show astrophysical variability.
Bottom row (Spectra): Arms show background spectra, Cosmic Ray events show hard spectra ($>2$~keV), and True Sources typically show softer spectra.
}
\label{fig:multimodal_comparison}
\end{figure*}

\end{enumerate}

\subsection{Pipeline Data Allocation Strategy}
To optimize the performance, we implemented a data allocation strategy:
\begin{itemize}
    \item Step 1 (Arm Filter): Utilizes only the Image Cutout ($X_{img}$).
    \item Step 2 (Cosmic Ray Filter): Combines Light Curve ($X_{lc}$) and Energy Spectrum ($X_{spec}$).
    \item Step 3 (Variability Screening): Reverts directly to the Screened Event File (\texttt{po\_cl.evt}).
\end{itemize}

\section{Methodology}
\label{sec:method}

\begin{figure}[t!]
    \centering
    \includegraphics[width=\textwidth]{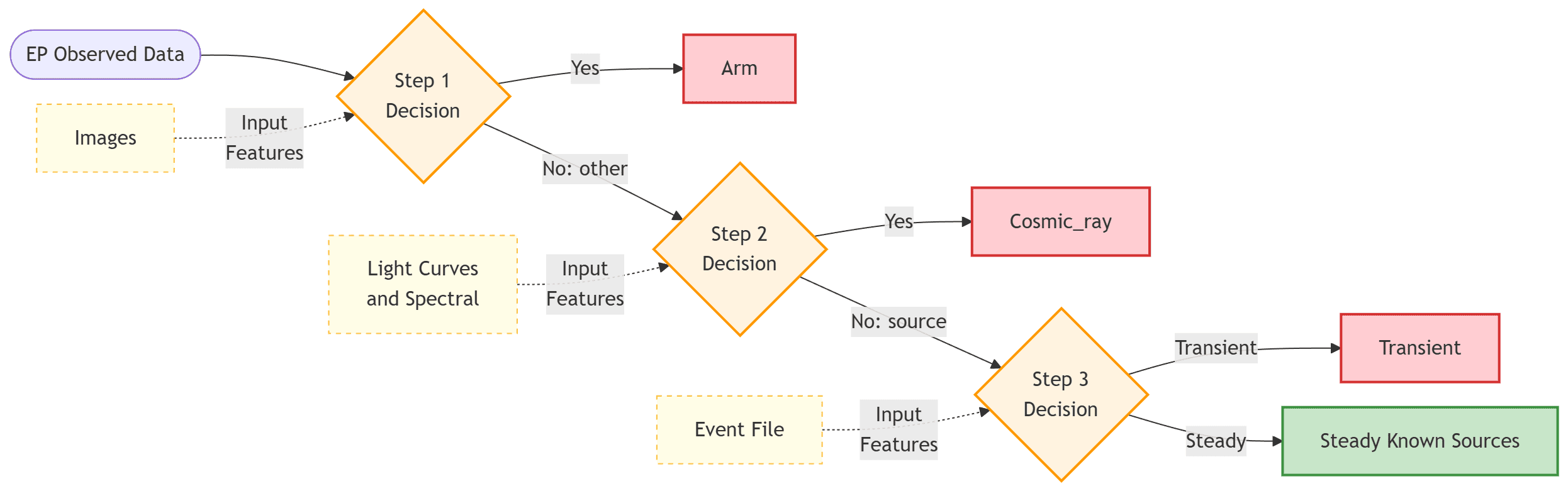} 
    \caption{\textbf{Overview of the M-EPDet Hierarchical Framework.}}
    \label{fig:framework}
\end{figure}

We propose a Multi-Step Hierarchical Classification Framework processing candidates sequentially through three steps: Arm Filter, Cosmic Ray Filter, and Variability Screening via Bayesian Blocks. The overall architecture is illustrated in Figure \ref{fig:framework}.

\subsection{Step 1: Arm Filter via ResNet}
\label{sec:method_step1}

\subsubsection{Physical Motivation and Task Definition}
The lobster-eye Micro-pore Optics (MPO) of EP-WXT produce a Point Spread Function (PSF) that differs from that of conventional focusing X-ray telescopes.
\begin{itemize}
    \item Genuine Sources: Bright astrophysical sources are physically characterized by a complete, symmetrical cruciform structure with a distinct central focal spot.
    \item Arm Artifacts: In contrast, the instrumental artifacts we aim to reject (referred to as ``Arms'') typically manifest as incomplete, asymmetrical, or isolated streak-like structures. These arise from ghost images, single reflections, or contamination from off-center sources.
\end{itemize}
Therefore, the objective of Step 1 is not to indiscriminately veto all cross-like shapes, but to discriminate morphological symmetry: distinguishing the isolated streaks of artifacts from the coherent cruciform PSF of real sources.

\subsubsection{Network Architecture and Feature Integration}
\begin{figure}[t!]
    \centering
    \includegraphics[width=\textwidth]{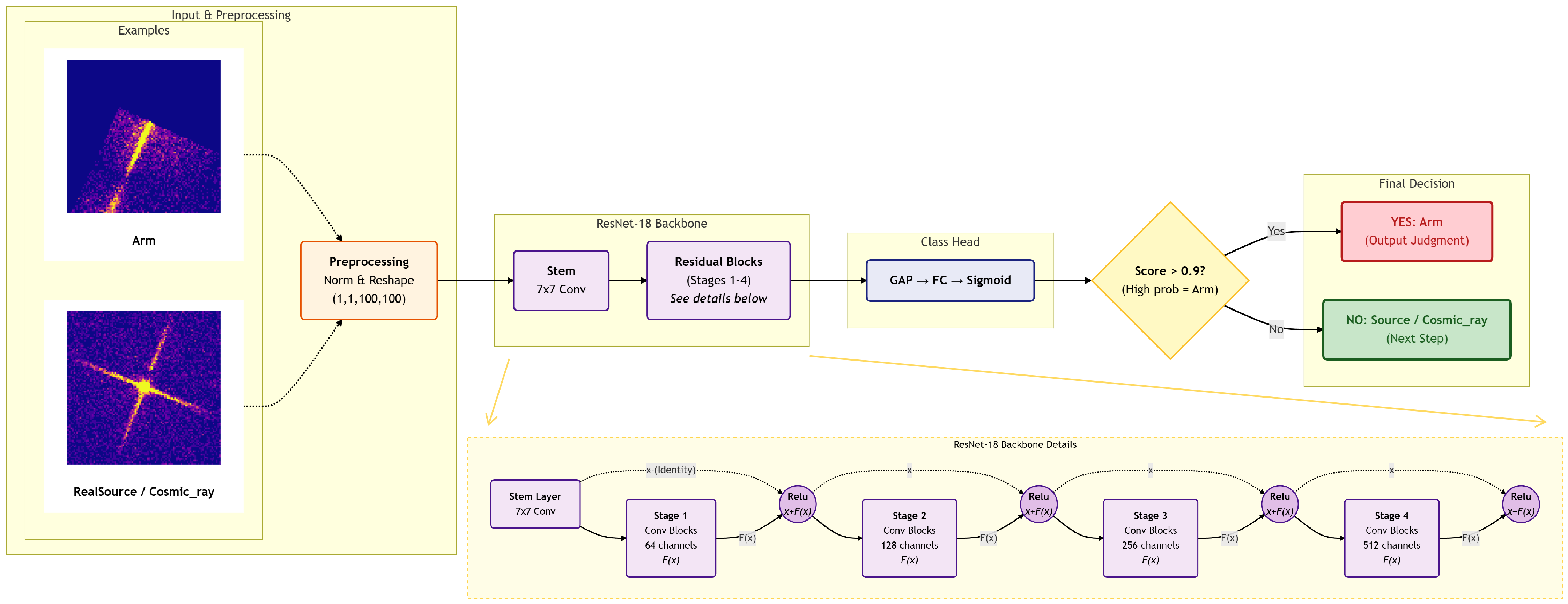} 
    \caption{\textbf{Architecture of the Step 1 Arm Filter.} 
    The pipeline takes preprocessed $100 \times 100$ image cutouts as input. 
    The ResNet-18 backbone utilizes four stages of residual blocks with explicit skip connections ($x + F(x)$) to fuse local textural features with global topological information. 
    The final decision logic applies a threshold ($P>0.9$) to explicitly VETO strong morphological artifacts (Arms) while passing potential candidates to the next step.}
    \label{fig:step1_combined}
\end{figure}

We adopt the ResNet-18 architecture \cite{He2016} as the backbone network, as illustrated in Figure \ref{fig:step1_combined}. The data flow proceeds from left to right: the left panel demonstrates the input preprocessing of candidate cutouts; the middle section visualizes the vertical stacking of residual blocks in the ResNet backbone; and the right panel depicts the simplified classification head that executes the final VETO logic based on the probability score.

In this task, residual connections are useful for two related reasons. Through hierarchical stacking and progressive downsampling, they expand the effective receptive field from local structures to the full $100 \times 100$ field, enabling the model to capture the global topology of elongated artifacts. At the same time, the identity mapping helps preserve shallow textural cues while deeper layers encode global morphology, improving discrimination between incomplete Arm-like streaks and coherent cruciform source structures.

\subsubsection{Input Preprocessing}
\label{sec:preprocess_step1}

The input consists of photon-count image cutouts ($100 \times 100$ pixels) centered on candidate sources. To adapt the raw astronomical images for the ResNet-18 model, we implemented the following preprocessing steps:

\begin{itemize}
    \item Instance-level Normalization: Given the significant dynamic range in exposure times and source fluxes across observations, we apply linear Min-Max Normalization independently to each image cutout. This maps the pixel intensities to the $[0, 1]$ interval:
    \begin{equation} 
        X_{norm} = \frac{X - X_{min}}{X_{max} - X_{min}}
    \end{equation}
    To ensure numerical stability, in edge cases where the image is constant (i.e., $X_{max} = X_{min}$), the pixel values are set to a zero matrix.
    
    \item Tensor Formatting: Each cutout is reshaped into a single-channel tensor for input to the ResNet-18 backbone.
\end{itemize}

\subsection{Step 2: Cosmic Ray Filter via Temporal and Spectral Features}
\label{sec:method_step2}

\begin{figure}[t!]
    \centering
    \includegraphics[width=\textwidth]{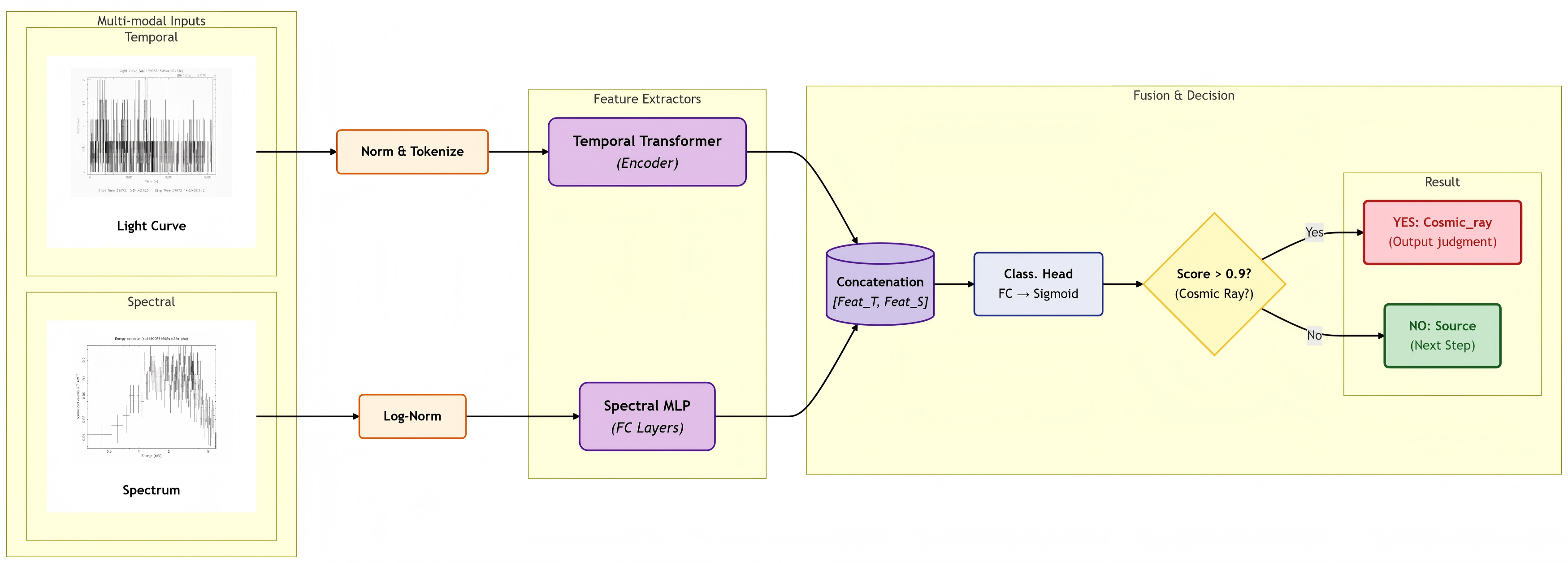}
    \caption{\textbf{Architecture of the Step 2 Cosmic Ray Filter.} To break the spatial degeneracy between Cosmic Ray events and genuine sources, a dual-branch network integrates multi-modal inputs. The temporal branch applies a Transformer encoder with an adaptive sliding window to the 3-channel light curve (time, rate, error). Concurrently, the spectral branch processes the 1D Pulse Invariant (PI) via an MLP. Globally pooled embeddings from both modalities are concatenated for the final classification.}
    \label{fig:step2_arch}
\end{figure}

To distinguish Cosmic Ray events from genuine astrophysical sources, we propose a dual-branch network architecture that simultaneously exploits temporal variability and spectral hardness. The overall framework is illustrated in Figure \ref{fig:step2_arch}.

\subsubsection{Physical Motivation: Breaking Spatial Degeneracy}
After spatial filtering, the remaining candidate pool is dominated by Cosmic Ray events mixed with genuine sources. In the spatial domain, these two classes are strongly degenerate morphologically, as both can appear point-like. However, they exhibit distinct signatures in other domains:
\begin{itemize}
    \item Temporal Domain: Cosmic Ray events are instantaneous events (Dirac delta function), whereas astrophysical transients typically exhibit resolvable rise and decay profiles.
    \item Spectral Domain: Cosmic Ray events deposit energy directly into the silicon, often producing a hard spectrum. Genuine focused sources are constrained to the Soft X-ray band (0.5--4 keV), exhibiting soft spectral characteristics.
\end{itemize}

\subsubsection{Feature Engineering and Importance Sampling}
To extract robust representations from sparse photon data while maintaining numerical stability for deep learning, we implemented a compact preprocessing pipeline:
\begin{enumerate}
    \item \textbf{Dual-Scale Normalization for Ternary Features}: We construct the light curve as a tensor of dimension $(L, 3)$, representing Time, Count Rate, and Poisson Error. The Time axis is normalized to $[0, 1]$ on an instance basis. To preserve physically informative count-rate differences across observations, we apply a logarithmic transformation, $X'_{count} = \log_{10}(X_{count} + 1)$, followed by a global Z-score standardization computed over the training corpus. The Poisson Error is scaled consistently with the count-rate channel.
    \item \textbf{Sliding Window Importance Sampling:} Because EP-WXT light curves have variable lengths and transient behavior may occupy only a small fraction of the observation, we select informative bins using a variance-based adaptive importance score. For a standardized count-rate sequence $\{\hat{r}_i\}$, the local score for each bin $i$ within a sliding window of size $w$ is defined as:
    \begin{equation}
        S_i = \mathrm{Var}\left(\hat{r}_{i-w/2}, \ldots, \hat{r}_{i+w/2}\right)
    \end{equation}
    The window size is dynamically set to $\max(5, \min(50, N/20))$ for a light curve with total length $N$. Bins with larger local variance are preferentially retained, and the selected bins are re-ordered chronologically. The corresponding physical time and Poisson error are concatenated with the count-rate channel so that the Transformer can retain the relevant temporal structure. The target sequence length $L$ is defined as the 95th percentile of the training-set length distribution and constrained to the range 200--3000 bins. For shorter sequences, we avoid interpolation in order to preserve the Poisson statistics of the photon-counting data; instead, we apply zero-padding together with a boolean attention mask that suppresses padded positions in self-attention by assigning them zero probability after softmax.
\end{enumerate}

\subsubsection{Dual-Branch Network Architecture}
We propose a Dual-Branch Network that extracts representations independently from the time and energy domains before performing a fused decision.
\begin{itemize}
    \item Temporal Branch (Transformer Encoder): We employ a Transformer Encoder \cite{Vaswani2017} injected with sinusoidal Positional Encodings. Leveraging Multi-head Self-Attention, the model captures long-range dependencies to distinguish between isolated impulses and continuous evolution profiles.
    \item Spectral Branch (MLP): This branch utilizes a Multi-Layer Perceptron (MLP) with GELU activations. It is designed to extract non-linear spectral shape features—implicitly learning the Hardness Ratio (HR)—to identify the characteristic hard spectral tail of Cosmic Ray.
    \item Feature Fusion: The vectors from both branches are concatenated and passed through a final classification head to output the probability.
\end{itemize}

\subsection{Step 3: Background-aware Variability Screening via Bayesian Blocks}
\label{sec:method_step3}

\subsubsection{Source and Background Event Extraction}
For each candidate, the source center is read from the corresponding DS9 region file. Events are extracted from a circular source aperture of radius $R_{\rm src}=67$ pixels, corresponding to approximately $9.2'$ ($\sim552''$) on the sky. 
This angular scale follows the source-extraction aperture adopted in the LEIA in-flight calibration analysis of \citet{cheng2025leia}, where a $9.2'$ circular source region was used to enclose more than 90\% of the focal-spot photons. Therefore, the adopted source aperture should be understood as an operational extraction region for the central focal spot, rather than as an aperture intended to enclose the full cruciform PSF including the extended Arms.

To estimate the local background, we construct a concentric annular region with inner and outer radii
\begin{equation}
R_{\rm in}=2R_{\rm src}, \qquad R_{\rm out}=4R_{\rm src}.
\end{equation}
These radii correspond to approximately $18.4'$ ($\sim1104''$) and $36.8'$ ($\sim2208''$), respectively, matching the background-annulus radii adopted by \citet{cheng2025leia}. The annulus provides a local background estimate around the candidate position while keeping the background region separated from the source aperture. Its area is 12 times the source-aperture area, providing improved counting statistics for the low-count variability screening in Step 3.

Only photons in the $0.5$--$4.0$~keV band are retained. Let $S$ and $B$ denote the source-region and background-region counts, respectively. The area-scaling factor is defined as
\begin{equation}
\alpha = \frac{R_{\rm out}^2 - R_{\rm in}^2}{R_{\rm src}^2},
\end{equation}
so that the background contribution expected inside the source aperture is estimated as $B/\alpha$. 
With the adopted annular geometry, $\alpha=12$.

\subsubsection{GTI Stitching and Background-aware Time Transformation}
When an observation contains multiple Good Time Intervals (GTIs), the event times are first stitched by removing the temporal gaps between adjacent GTIs. The subsequent analysis is therefore performed on the effective exposure timeline rather than on the raw wall-clock axis.

Bayesian Blocks is not applied directly to the raw source-region arrival times. Instead, let $\{t_i^{\rm src}\}$ and $\{t_j^{\rm bkg}\}$ be the stitched source and background event times. We map each source event onto a background-referenced transformed axis through the empirical cumulative distribution of the background events:
\begin{equation}
u_i=\frac{\#\{t_j^{\rm bkg}\le t_i^{\rm src}\}}{N_{\rm bkg}}.
\end{equation}
If the source-region events follow the same temporal distribution as the local background, the transformed sequence $\{u_i\}$ is expected to be approximately uniform. Deviations from uniformity therefore indicate source variability relative to the local background rather than fluctuations of the background itself.

\subsubsection{Bayesian Blocks Segmentation and Dynamic Calibration}
We apply the Bayesian Blocks algorithm for time-tagged event data \cite{Scargle2013} to the transformed event sequence $\{u_i\}$. The segmentation is controlled by the change-point penalty $\gamma$ (equivalently, \texttt{ncp\_prior}), which determines the cost of introducing a new block. We adopt the empirical calibration of \citet{Scargle2013},
\begin{equation}
    \gamma(N, p_0) = 4 - \ln(73.53 \cdot p_0 \cdot N^{-0.478}),
    \label{eq:ncp_prior}
\end{equation}
where $p_0$ is the target false positive rate and $N$ is the number of source-region photons. In the implementation, the corresponding \texttt{ncp\_prior} values are precomputed from this expression and stored in a lookup table for efficient batch processing.

After segmentation on the transformed axis, the block edges are mapped back to physical time using the empirical percentiles of the stitched background-event distribution. This yields a set of time-domain blocks over which source, background, and net count rates can be estimated consistently.

\subsubsection{Block-level Screening Criteria}
After the Bayesian Blocks segmentation is mapped back to the physical time axis, we compute the source counts, background counts, background-scaled count rates, and background-subtracted net count rates for each block. Variability significance is then evaluated from the change in background-subtracted net count rate between adjacent blocks, with the uncertainty propagated from the source and background counting statistics.

A candidate is retained by Step 3 only if all of the following conditions are satisfied:
\begin{enumerate}
    \item the Bayesian Blocks segmentation yields at least two blocks ($N_{\rm blocks}\ge2$);
    \item the maximum adjacent-block significance of the background-subtracted net count-rate change exceeds the adopted threshold ($\Delta \sigma_{\max}\ge6$ in the current implementation);
    \item the source region contains at least 20 photons ($N_{\rm src}\ge20$).
\end{enumerate}

These criteria suppress spurious triggers driven by local background fluctuations and low-count Poisson noise, while retaining observations with statistically significant background-corrected intra-observation variability.

\section{Results}
\label{sec:results}

In this chapter, we evaluate the performance of the three proposed processing steps on a standard validation dataset. We present both the foundational metrics and comparative experiments against baseline methods.

\subsection{Experimental Setup}
\label{subsec:exp_setup}

In this section, we detail the experimental configurations to ensure the reproducibility of our results. We describe how the dataset was partitioned, the specific hyperparameters used for model training, and the metrics adopted for performance evaluation.

\subsubsection{Dataset Partitioning}
To assess generalization under a random split, the labeled dataset constructed in Section \ref{sec:data} was randomly partitioned into three subsets with a ratio of \textbf{8:1:1}:
\begin{itemize}
    \item \textbf{Training Set (80\%):} Used for model parameter optimization via backpropagation.
    \item \textbf{Validation Set (10\%):} The validation set was used for hyperparameter tuning, model selection, and, where applicable, for early stopping and scheduler control.
    \item \textbf{Test Set (10\%):} Used exclusively for the final performance evaluation reported in this section. This subset was never seen by the model during the training phase to ensure an unbiased evaluation.
\end{itemize}

The split was performed at the candidate-observation level rather than at the unique-source level. Consequently, repeated observations of the same astrophysical source may appear in the training, validation, and test subsets. This setting is consistent with the operational EP-WXT candidate stream, where the pipeline processes and vets each candidate observation independently. Therefore, this evaluation should be interpreted as an observation-level operational benchmark rather than a strictly source-disjoint generalization test.

Because the raw observational stream is extremely imbalanced, we constructed step-specific balanced binary benchmarks for Steps 1 and 2. For Step 1, the benchmark is defined as Arm versus Other, where the Other class consists of Source and Cosmic Ray samples; to avoid the Other class being dominated by genuine sources, Source and Cosmic Ray samples were first balanced within the Other class before constructing the binary benchmark against the Arm samples. For Step 2, we retained all available Cosmic Ray samples and randomly down-sampled the Source class to the matched sample size.

\subsubsection{Implementation Details and Hyperparameters}
All deep learning models in this work were implemented using the \textbf{PyTorch} framework. 
The experiments were conducted on a workstation equipped with an NVIDIA A100 GPU (40GB VRAM). Mixed Precision (AMP) was used where applicable.

Due to their distinct data modalities, we employed different training strategies for the Arm Filter (Step 1) and the Cosmic Ray Filter (Step 2). The network architectures, optimization parameters, and data augmentation methods are detailed in Table~\ref{tab:hyperparameters}.

\begin{table*}[t]
\centering
\caption{Detailed Hyperparameters and Architectural Configurations for M-EPDet}
\label{tab:hyperparameters}
\resizebox{\textwidth}{!}{
\begin{tabular}{@{}ll|ll@{}}
\toprule
\multicolumn{2}{c|}{\textbf{Step 1: Arm Filter (ResNet-18)}} & \multicolumn{2}{c}{\textbf{Step 2: Cosmic Ray Filter (Transformer + MLP)}} \\ \midrule
\multicolumn{2}{l|}{\textit{A. Model Architecture}} & \multicolumn{2}{l}{\textit{A. Model Architecture}} \\
Input Modification & Conv2d(1, 64, kernel=7, stride=2, padding=3) & Input Features & Time, Rate (Log), Error (Log) \\
Input Dimensions & $100 \times 100$ pixels (Single Channel) & Max Sequence Length & $3000$ (Temporal), $1024$ (Spectral PI) \\
Pooling Strategy & AdaptiveAvgPool2d((1, 1)) & Transformer Layers & $6$ Layers, $8$ Attention Heads \\
Output Dimension & 2 (Arm vs. Other) & Hidden Dimensions & $d_{model} = 512$, Feedforward = $2048$ \\
Dropout Rate & $0.5$ (Before FC Layer) & Spectral Embedding & Dimension = $64$ \\
Total Parameters & $\sim 11.2$ Million & Pooling Strategy & 1D Global Average Pooling \\ \midrule
\multicolumn{2}{l|}{\textit{B. Training \& Optimization}} & \multicolumn{2}{l}{\textit{B. Training \& Optimization}} \\
Optimizer & Adam & Optimizer & AdamW ($\beta_1=0.9, \beta_2=0.999$) \\
Learning Rate (LR) & $1 \times 10^{-4}$ & Learning Rate (LR) & $5 \times 10^{-5}$ (with CosineAnnealingLR) \\
Weight Decay & $1 \times 10^{-4}$ & Weight Decay & $0.01$ \\
LR Scheduler & ReduceLROnPlateau (factor=0.5, patience=5) & Warmup Steps & $1000$ \\
Batch Size / Epochs & $64$ / $30$ & Batch Size / Epochs & $64$ / $10$ (with AMP) \\
Early Stopping & Patience = 10 & Early Stopping & N/A \\
Loss Function & Weighted Cross-Entropy Loss & Loss Function & Cross-Entropy Loss \\
Gradient Clipping & N/A & Gradient Clipping & Max Norm = $1.0$ \\ \midrule
\multicolumn{2}{l|}{\textit{C. Data Augmentation}} & \multicolumn{2}{l}{\textit{C. Data Augmentation}} \\
Random Rotation & $\pm 15^\circ$ & Gaussian Noise & Std = $0.01$ \\
Random Flip & Horizontal (50\%) \& Vertical (50\%) & Time Shift & Max 100 steps \\
Gaussian Blur & $30\%$ prob (kernel=3, $\sigma \in [0.1, 2.0]$) & Amplitude Scale & Random scaling $\in [0.8, 1.2]$ \\
Importance Sampling & N/A & Target Percentile & $95\%$ (Min length truncation = 200) \\ \bottomrule
\end{tabular}
}
\end{table*}

The Step 2 model was trained for 10 epochs, which was empirically sufficient for stable validation performance.

\subsubsection{Evaluation Metrics}
To quantitatively evaluate the classification performance, we utilize standard metrics derived from the confusion matrix. In the context of our cascading ``Real-Bogus'' classification task, the positive class is defined in a step-specific manner. For Step 1 (Arm Filter), the positive class is \textbf{Other} (i.e., Source + Cosmic Ray), and the negative class is \textbf{Arm}. For Step 2 (Cosmic Ray Filter), the positive class is \textbf{Source}, and the negative class is \textbf{Cosmic Ray}.

Based on the confusion-matrix entries, we report \textbf{Precision}, \textbf{Recall}, and the \textbf{F1-score}:
\begin{equation}
    \text{Precision} = \frac{TP}{TP + FP}, \quad
    \text{Recall} = \frac{TP}{TP + FN}, \quad
    \text{F1} = 2 \cdot \frac{\text{Precision} \cdot \text{Recall}}{\text{Precision} + \text{Recall}}
\end{equation}

In our evaluation, we prioritize Recall as the primary metric. Because the scientific penalty of missing a rare transient (False Negative) significantly outweighs the operational cost of reviewing residual false alarms (False Positive), we adopt a conservative veto threshold of 0.9 in operation. Furthermore, given the large held-out test set sizes for Steps 1 and 2, we report point estimates strictly at this adopted operating threshold. For Step 1 and Step 2, we additionally report the ROC-AUC and PR-AUC of the positive class (i.e., Other for Step 1 and Source for Step 2) to characterize the threshold-independent ranking performance of the models.

\subsection{Performance of Step 1: Arm Filter}
\label{subsec:exp_step1}

\begin{figure}[h]
    \centering
    \begin{subfigure}[b]{0.48\textwidth}
        \centering
        \includegraphics[width=\textwidth]{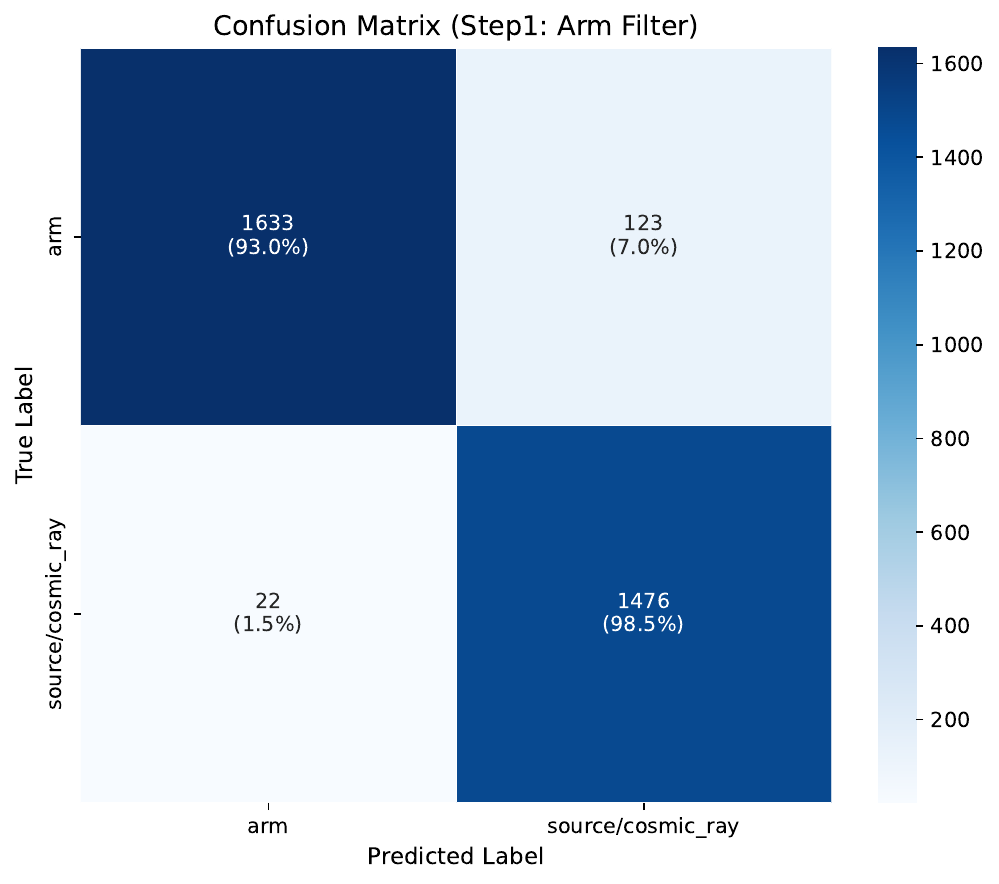}
        \caption{Confusion Matrix}
    \end{subfigure}
    \hfill
    \begin{subfigure}[b]{0.48\textwidth}
        \centering
        \includegraphics[width=\textwidth]{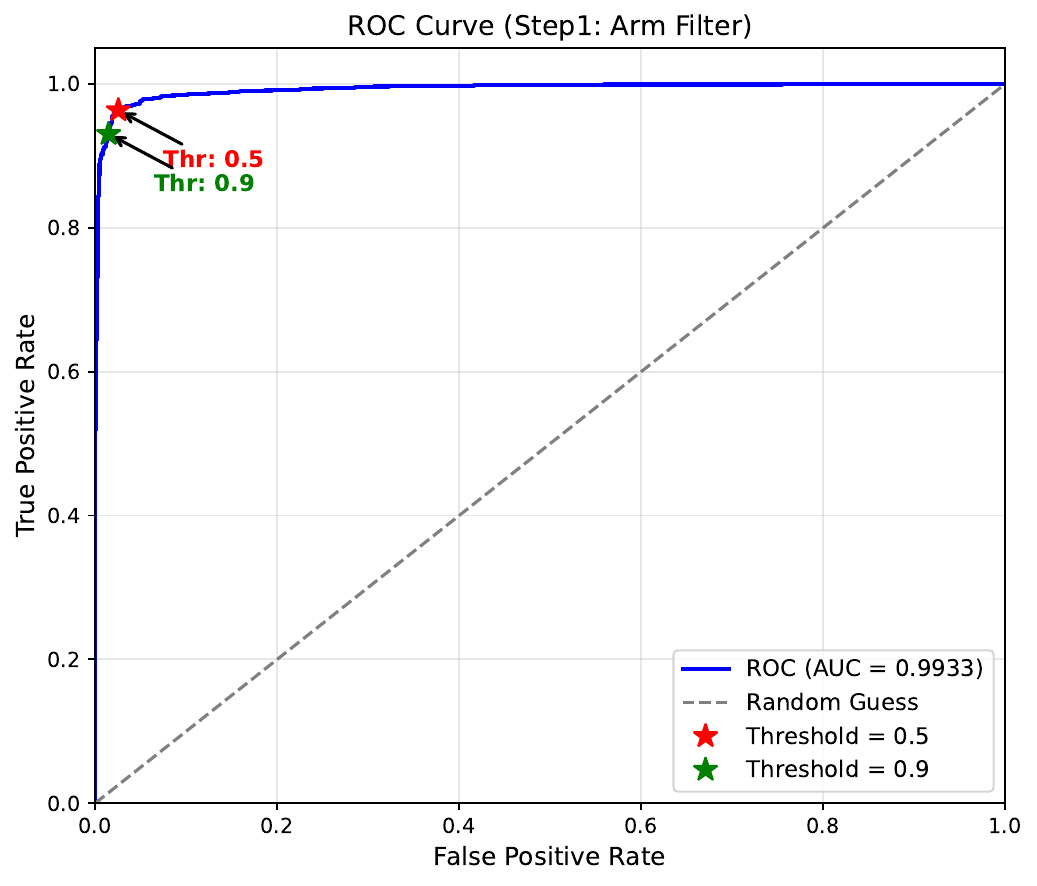}
        \caption{ROC Curve}
    \end{subfigure}
    \caption{\textbf{Performance visualization of Step 1 on the test set.} The confusion matrix and ROC curve jointly summarize the classification behavior of the ResNet-18 Arm Filter.}
    \label{fig:step1_cm}
\end{figure}

The ResNet-18 Arm Filter achieved an overall Accuracy of 95.54\% on the independent test set.
Table~\ref{tab:step1_performance} summarizes the metrics. The model achieves a Recall of 98.53\% for the Other (Source/Cosmic Ray) class, indicating that most genuine transient candidates are retained at the adopted threshold. For the positive class, the model further achieved an ROC-AUC of 0.9933 and a PR-AUC of 0.9915, indicating good threshold-independent ranking performance. Figure \ref{fig:step1_cm} presents the corresponding confusion matrix and ROC curve.

\begin{table}[h]
    \centering
    \caption{\textbf{Performance metrics of the Step 1 Arm Filter.}}
    \label{tab:step1_performance}
    \begin{tabular}{lccc}
        \toprule
        \textbf{Class} & \textbf{Precision (\%)} & \textbf{Recall (\%)} & \textbf{F1-Score (\%)} \\
        \midrule
        Other (Source/Cosmic Ray) & 92.30 & 98.53 & 95.31 \\
        Arm & 98.67 & 92.99 & 95.74 \\ 
        \bottomrule
    \end{tabular}
\end{table}

\subsubsection{Comparison with Baseline Methods}
To evaluate the relative performance of the ResNet-18 architecture, we benchmarked it against traditional machine learning and shallow deep learning baselines. 

\begin{figure}[htbp]
    \centering
    \includegraphics[width=0.7\textwidth]{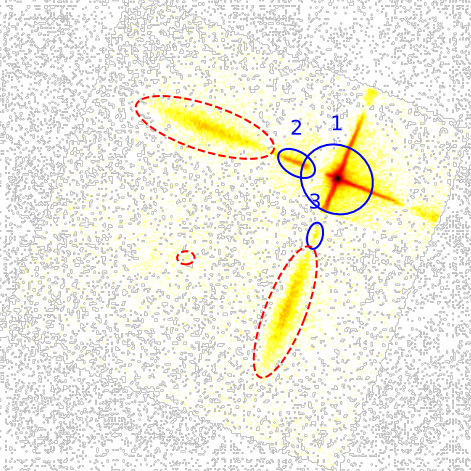}
    \caption{\textbf{Failure case of traditional source extraction on EP-WXT data.} The overlaid markers represent the classification results from SExtractor: red regions are identified as Arm artifacts, while blue circles indicate presumed point sources. Notably, the region highlighted by \textbf{blue circle \#2} is physically a discontinuous segment of an extended Arm artifact. Because traditional contour-based extractors fail to recognize the broken Arm as a single contiguous entity, this fragment is erroneously isolated and misclassified as a genuine source. This structural fragmentation renders parametric morphological metrics (e.g., ellipticity) ineffective, necessitating representation-learning approaches.}
    \label{fig:sextractor_failure}
\end{figure}

Prior to deep learning, traditional astronomical pipelines heavily relied on source extraction software (e.g., SExtractor) to veto artifacts based on morphological parameters such as ellipticity. However, initial exploratory tests suggested that such parametric methods are not well suited to the EP-WXT data considered here. The cruciform Arm artifacts are often discontinuous and mottled in low-SNR environments. As visually demonstrated in Figure~\ref{fig:sextractor_failure}, traditional contour-based extractors fail to recognize the extended Arm as a single contiguous entity, instead fragmenting it into multiple spurious point-like sources. This structural fragmentation weakens the usefulness of global morphological metrics such as ellipticity, making direct quantitative comparison difficult. 

Therefore, our quantitative benchmark focuses on feature-extraction and representation-learning methods evaluated under the same operating condition. For a fair comparison, all methods were tested on the identical held-out split, and the final class assignment for each model was reported under the same conservative veto threshold of 0.9. The results are summarized in Table~\ref{tab:step1_comparison}.

Among the compared baselines, HOG+SVM \cite{Cortes1995} performs substantially worse than the neural-network models, achieving 63.58\% Accuracy and an Arm Recall of 33.49\%, suggesting that hand-crafted HOG features are less effective for Arm rejection under the adopted conservative veto criterion. The shallow convolutional baselines, SimpleCNN and LeNet-5 \cite{LeCun1998}, perform substantially better, with Accuracies of 92.96\% and 93.36\%, respectively, but still remain inferior to ResNet-18.

Overall, ResNet-18 provides the most favorable trade-off among the methods evaluated here between candidate preservation and artifact suppression. It achieves the highest Accuracy (95.54\%), the highest Arm Recall (92.99\%), and the best ranking performance among the methods evaluated here (ROC-AUC = 0.9933, PR-AUC = 0.9915). Although SimpleCNN yields a slightly higher Recall for the Other class, this comes at the cost of weaker Arm rejection. Taken together, these results suggest that ResNet-18 offers a favorable overall trade-off between preserving scientifically relevant candidates and suppressing instrumental Arm artifacts under the adopted conservative veto threshold.

\begin{table*}[htbp]
\centering
\caption{\textbf{Full comparison of Step 1 baseline methods under the same conservative veto threshold of 0.9.} Metrics are reported for both classes. PR-AUC is reported for the Other (Other = Source + Cosmic Ray.) class, which is the positive class in Step 1.}
\label{tab:step1_comparison}
\resizebox{\textwidth}{!}{
\begin{tabular}{lccccccccc}
\hline
Method & Accuracy & P(Other) & R(Other) & F1(Other) & P(Arm) & R(Arm) & F1(Arm) & ROC-AUC & PR-AUC (Other) \\
\hline
HOG+SVM   & 63.58\% & 55.91\% & 97.87\% & 71.17\% & 97.19\% & 33.49\% & 49.81\% & 0.8552 & 0.8240 \\
SimpleCNN & 92.96\% & 87.43\% & 98.73\% & 92.74\% & 98.97\% & 87.87\% & 93.09\% & 0.9905 & 0.9864 \\
LeNet-5   & 93.36\% & 88.43\% & 98.46\% & 93.18\% & 98.55\% & 89.01\% & 93.54\% & 0.9876 & 0.9818 \\
ResNet-18 & 95.54\% & 92.30\% & 98.53\% & 95.31\% & 98.67\% & 92.99\% & 95.74\% & 0.9933 & 0.9915 \\
\hline
\end{tabular}
}
\end{table*}

\subsection{Performance of Step 2: Cosmic Ray Filter}
\label{subsec:exp_step2}

\begin{figure}[h]
    \centering
    \begin{subfigure}[b]{0.48\textwidth}
        \centering
        \includegraphics[width=\textwidth]{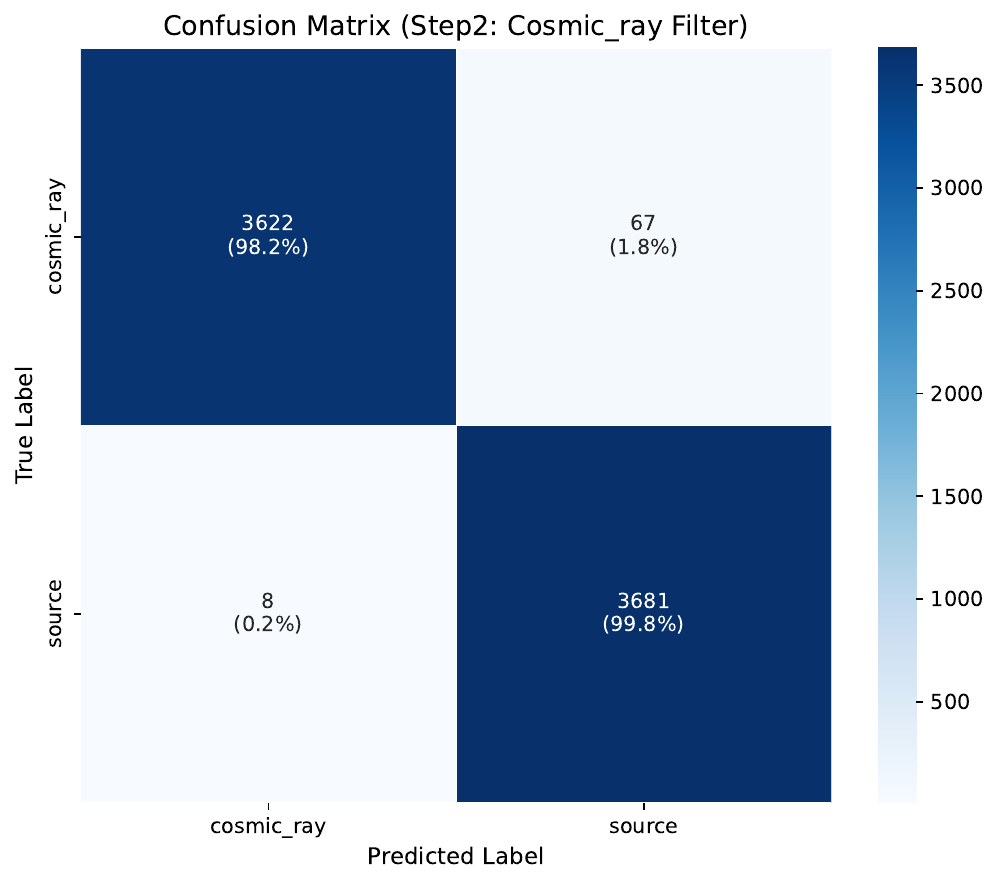}
        \caption{Confusion Matrix}
    \end{subfigure}
    \hfill
    \begin{subfigure}[b]{0.48\textwidth}
        \centering
        \includegraphics[width=\textwidth]{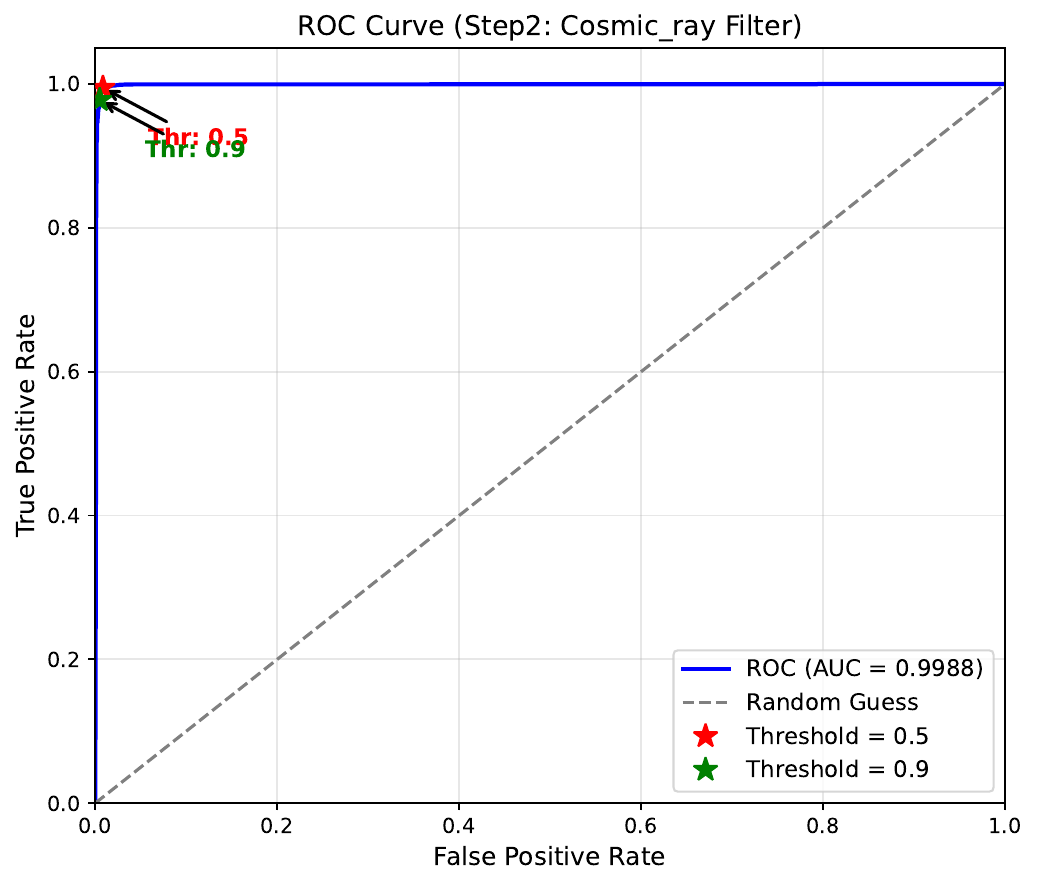}
        \caption{ROC Curve}
    \end{subfigure}
    \caption{\textbf{Performance visualization of Step 2 on the test set.} The confusion matrix and ROC curve jointly summarize the classification behavior of the dual-branch Cosmic Ray Filter.}
    \label{fig:step2_cm}
\end{figure}

The dual-branch Cosmic Ray Filter achieved an overall Accuracy of 98.98\% on the independent test set. Table~\ref{tab:step2_performance} summarizes the metrics. The model achieves a Recall of 99.78\% for the Source class, indicating that most genuine transient candidates are retained at the adopted threshold. For the positive class, the model further achieved an ROC-AUC of 0.9988 and a PR-AUC of 0.9979, indicating good threshold-independent ranking performance. Figure \ref{fig:step2_cm} shows the Confusion Matrix and ROC curve.

\begin{table}[h]
    \centering
    \caption{\textbf{Performance metrics of the Step 2 Cosmic Ray Filter.}}
    \label{tab:step2_performance}
    \begin{tabular}{lccc}
        \toprule
        \textbf{Class} & \textbf{Precision (\%)} & \textbf{Recall (\%)} & \textbf{F1-Score (\%)}\\
        \midrule
        Source & 98.21 & 99.78 & 98.99 \\
        Cosmic Ray  & 99.78 & 98.18 & 98.97 \\
        \bottomrule
    \end{tabular}
\end{table}

\subsubsection{Comparison with Traditional Machine Learning Approaches}
Existing pipelines often employ traditional machine learning algorithms (e.g., Random Forest) that rely on hand-crafted features for Cosmic Ray rejection \cite{Zuo2024}. While such methods can perform well for standard Cosmic Ray, they may face a physical ambiguity when evaluating temporally compact astrophysical candidates. In binned light curves, short-timescale transient candidates or sparsely sampled burst-like events can appear as isolated spike-like features, making them partially degenerate with Cosmic Ray in the temporal domain. A classifier relying primarily on temporal morphology may therefore risk vetoing such candidates as Cosmic Ray events. To reduce this risk, M-EPDet integrates temporal information with the calibrated energy-distribution information from the 1D Pulse Invariant (PI). The spectral branch provides an additional discriminant between particle-induced events and focused X-ray photons, thereby improving Cosmic Ray rejection while reducing the risk of misclassifying short-timescale transient candidates. To quantify the contribution of the spectral branch beyond individual examples, we performed an additional ablation experiment on an independent held-out test set sampled from observations obtained after March 2026. This independent set does not overlap with the training data used in this work, which were collected between July 10, 2024 and July 10, 2025. The test set initially contained 1,000 astrophysical sources and 1,000 Cosmic Ray samples. After excluding seven samples with missing or empty files, 1,993 valid samples remained, including 996 sources and 997 Cosmic Ray samples. Using the same threshold criterion of 0.9 adopted in the main evaluation, the light-curve-only Transformer achieved an accuracy of 93.68\%, whereas the dual-branch temporal-spectral model achieved 99.40\%. The number of misclassified samples decreased from 126 to 12. The dual-branch model corrected 119 samples misclassified by the light-curve-only model. The results are summarized in Table~\ref{tab:step2_ablation_independent}. These results provide quantitative evidence that the spectral branch supplies complementary information and substantially improves the separation between genuine astrophysical sources and Cosmic Ray events.

\begin{table}
\centering
\caption{Ablation evaluation of the Step 2 Cosmic Ray Filter on an independent held-out test set obtained after March 2026. The same threshold criterion of 0.9 adopted in the main evaluation is used for both models.}
\label{tab:step2_ablation_independent}
\begin{tabular}{lcccc}
\toprule
Model & Accuracy & Source Recall & Cosmic Ray Recall & Misclassified \\
\midrule
Light-curve-only Transformer & 93.68\% & 96.39\% & 90.97\% & 126 \\
Dual-branch temporal-spectral model & 99.40\% & 99.60\% & 99.20\% & 12 \\
\bottomrule
\end{tabular}
\end{table}

\begin{figure*}[htbp]
\centering
\gridline{
    \fig{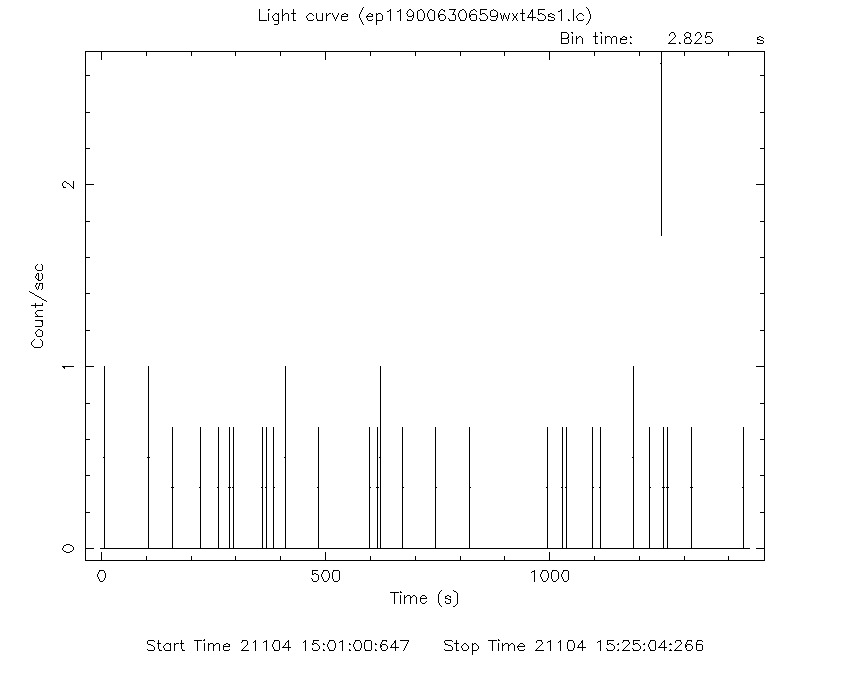}{0.32\textwidth}{(a) Case 1: Light curve}
    \fig{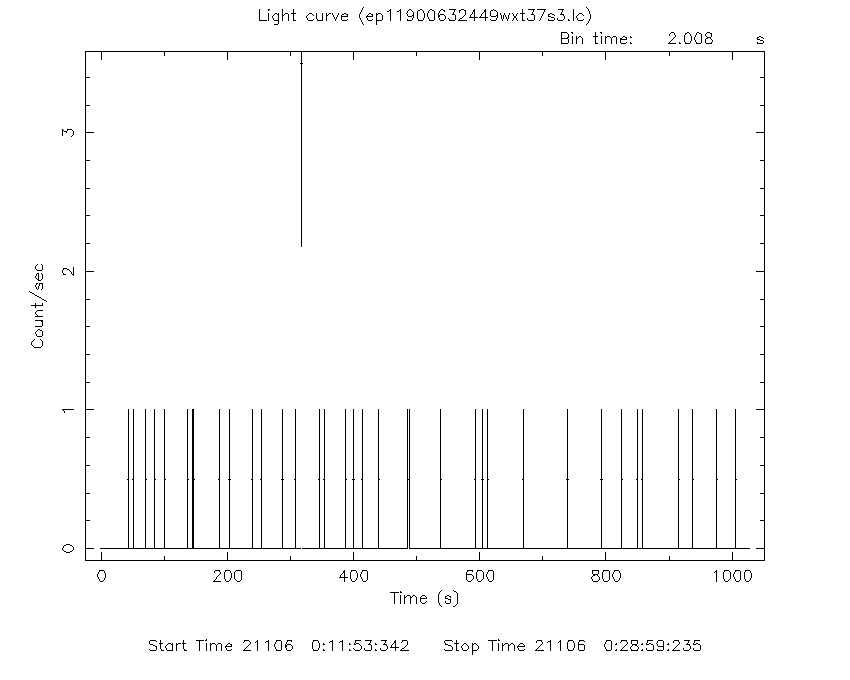}{0.32\textwidth}{(b) Case 2: Light curve}
    \fig{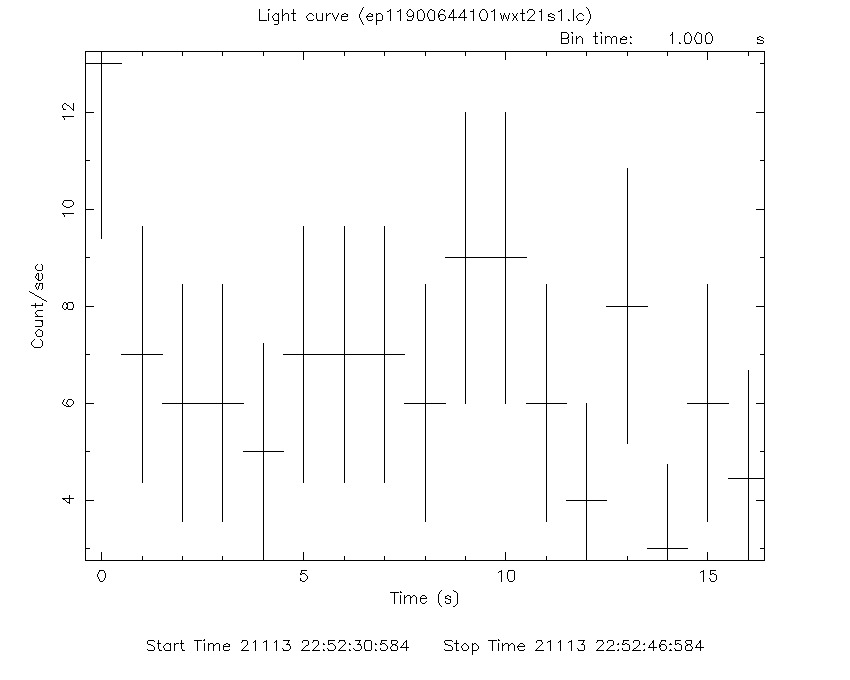}{0.32\textwidth}{(c) \texttt{ep11900644101wxt21s1}: Light curve}
}
\gridline{
    \fig{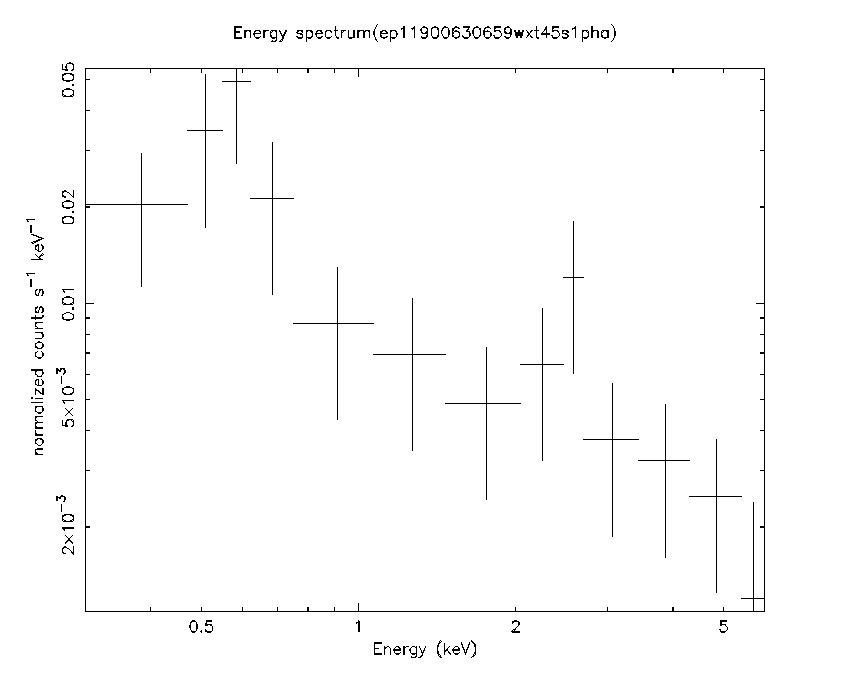}{0.32\textwidth}{(d) Case 1: PI spectrum}
    \fig{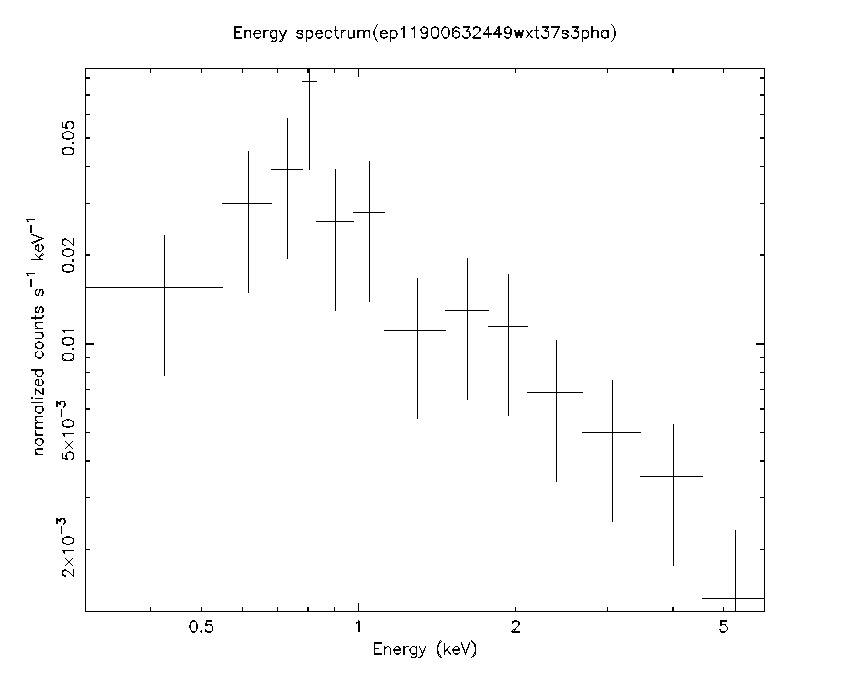}{0.32\textwidth}{(e) Case 2: PI spectrum}
    \fig{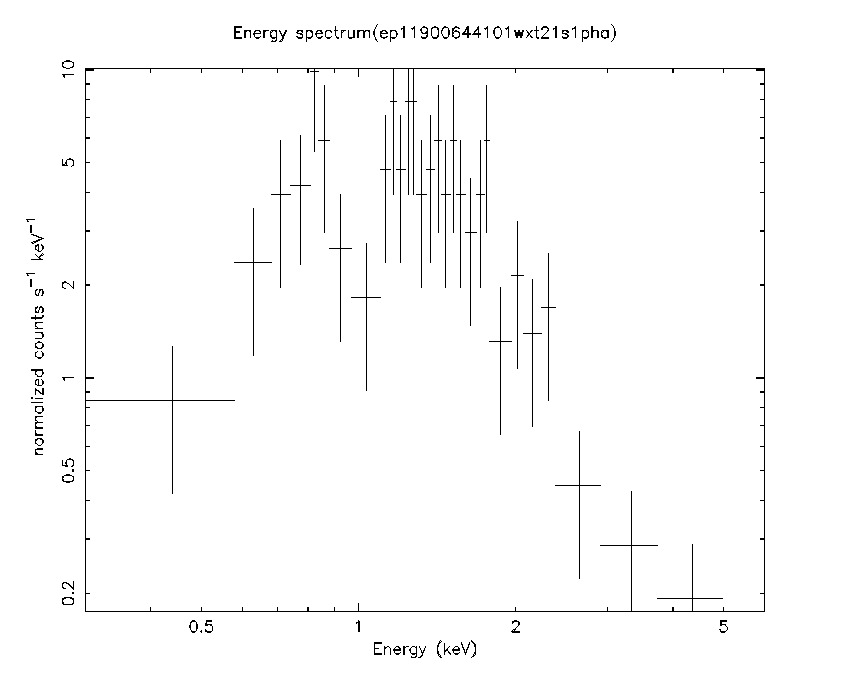}{0.32\textwidth}{(f) \texttt{ep11900644101wxt21s1}: PI spectrum}
}
\caption{
Light curves and PI spectra of the three verified astrophysical candidates used as illustrative examples for the Step 2 spectral-branch analysis}. The first two cases illustrate temporally compact or spike-like source morphology, while the third case (\texttt{ep11900644101wxt21s1}) is included as an operational genuine-source example whose Temporal-Only misclassification was corrected by the spectral branch.

\label{fig:step2_ablation_cases}
\end{figure*}

In addition to this quantitative ablation, we retain three verified astrophysical candidates from the operational EP-WXT pipeline as illustrative examples of the practical effect of the spectral branch on candidate selection. The corresponding light curves and PI spectra are shown in Figure \ref{fig:step2_ablation_cases}. Under a Temporal-Only baseline, all three cases received Cosmic Ray probabilities above 90\% and would therefore have been rejected. With the full dual-branch model, the additional PI information reduced the Cosmic Ray probability to about 74\% for two cases and yielded a genuine-source probability of 98\% for the third. One of these retained cases, \texttt{ep11900644101wxt21s1}, corresponds to the fast X-ray transient EP260314a later reported in \cite{GCN44016}. We emphasize that this is an illustrative operational case study rather than a statistical evaluation; a broader assessment will require a larger sample of confirmed FXT events.

\subsection{Performance of Step 3: Variability Screening via Bayesian Blocks}
\label{subsec:exp_step3}

The final step is designed as a single-exposure variability screening module rather than a source-type classifier. Its purpose is to extract a compact subset of observations exhibiting statistically significant intra-observation flux changes from a much larger historical archive, thereby helping reduce the manual burden of downstream scientific vetting. To quantify this screening behavior, we applied Step 3 to all observations that survived Steps 1 and 2 between August 10, 2024 and March 9, 2026. The input pool comprised 282,099 candidate observations, from which the background-aware Bayesian Blocks module retained 2,117 for further inspection.

\subsubsection{Archive-scale Candidate Reduction}
Relative to the 282,099 observations entering Step 3, the retained set of 2,117 observations corresponds to a screening retention rate of 0.75\% and a candidate-volume reduction of 99.25\%. This degree of compression is operationally useful: it reduces an archive-scale observation stream to a subset that is more manageable for manual scientific inspection, while concentrating attention on observations with short-timescale variability signatures.

\subsubsection{Composition of the Retained Subset}
The retained subset is not uniformly populated across source populations. Among the 2,117 flagged observations, 1,605 are associated with X-ray binaries (XRBs), 23 with GRBs, and 76 with transient sources. In fractional terms, these correspond to 75.81\%, 1.09\%, and 3.59\% of the retained sample, respectively.

Because X-ray binaries are frequently revisited and can contribute multiple retained observations from the same object, we performed a secondary population-level analysis after grouping by \texttt{simbad\_name}. Restricting to sources with fewer than five occurrences yields 517 observations, among which 85 are associated with XRBs, 23 with GRBs, and 63 with transient sources, corresponding to 16.44\%, 4.45\%, and 12.19\%, respectively. The decrease in the transient category from 76 to 63 after applying the \texttt{simbad\_name} occurrence cut is due to repeated detections of the same transient source names in the full retained sample, rather than to the removal of 13 distinct transient sources. In this secondary analysis, source names with five or more occurrences are excluded by construction. The GRB count remains unchanged at 23 because the GRB-associated source names do not exceed the same occurrence threshold. Relative to the full retained set, the GRB fraction therefore rises from 1.09\% to 4.45\%, and the transient fraction from 3.59\% to 12.19\%; if GRB- and transient-tagged observations are combined, their joint fraction increases from 4.68\% to 16.63\%. Moreover, 58 of the 63 transient-tagged observations (92.1\%) in this secondary subset carry EP-style source names (e.g., EPYYMMDDa), suggesting that the retained sample is enriched not only in relatively rare outburst-like events but also in recently designated EP transients after repeated-source suppression. In practical operation, the module often flags explosive phenomena such as Type I X-ray bursts and fast X-ray transients, which is consistent with the intended role of Step 3 as a rapid-variability-driven screening step for scientific follow-up.

\begin{table}[h]
    \centering
    \caption{\textbf{Archive-scale screening statistics of Step 3.} The Bayesian Blocks module was applied to all 282,099 observations that survived Steps 1 and 2 between 2024 August 10 and 2026 March 9. The table reports the composition of the 2,117 retained observations, together with a secondary count after restricting to sources with fewer than five occurrences of the same simbad\_name.}
    \label{tab:step3_archive_stats}
    \begin{tabular}{lcccc}
        \toprule
        \textbf{Category} & \textbf{Flagged} & \textbf{Fraction in 2,117} & \textbf{After source-occurrence cut} & \textbf{Fraction in 517} \\
        \midrule
        XRB       & 1605 & 75.81\% & 85 & 16.44\% \\
        GRB       & 23   & 1.09\%  & 23 & 4.45\% \\
        Transient & 76   & 3.59\%  & 63 & 12.19\% \\
        \bottomrule
    \end{tabular}
\end{table}

Taken together, these results indicate that Step 3 functions as an effective variability-driven screening step: it filters out most weakly varying observations, while yielding a compact subset enriched in relatively rare outburst phenomena. This behavior directly supports its operational purpose in the EP-WXT pipeline, namely, reducing the human vetting load by concentrating attention on observations with statistically significant variability signatures.

\subsubsection{Method Selection}
Prior to adopting the Bayesian Blocks algorithm, we explored deep learning-based approaches for this screening problem, including 1D-CNNs, LSTMs, and Transformer-based architectures. However, extensive experiments showed that these supervised sequence models failed to learn stable and transferable representations in the low-count EP-WXT regime, with validation accuracy remaining near chance level.

This result is consistent with the statistical properties of faint EP-WXT event data. In this regime, the relevant signal is whether an observation shows statistically significant variability relative to the local background, rather than whether it contains a stable learned temporal morphology. Bayesian Blocks is therefore better suited to this task because it directly tests for significant change points in sparse event data.

\section{Application and Discussion}
\label{sec:application}

In this chapter, we present the deployment details of the M-EPDet pipeline within the EP-WXT data processing pipeline \cite{zhang2025} and evaluate its on-orbit performance. Furthermore, we discuss critical factors influencing model accuracy—specifically the impact of image crop size—and outline our preliminary exploration into fine-grained source classification using physical parameters.

\subsection{Operational Deployment and On-orbit Performance}
\label{subsec:deployment}

\subsubsection{System Architecture and Efficiency}
The proposed framework has been integrated into the EP-WXT data processing pipeline. To ensure portability and ease of maintenance, the cascading classification model is encapsulated within a Docker container. 
Notably, unlike the training phase which requires high-performance GPUs, the inference module is deployed on CPU-only servers in the production environment.
Despite the absence of hardware acceleration, the relatively lightweight design of our hierarchical models (ResNet-18 and compact Transformers) allows efficient inference. Operational logs show that the system processes a single candidate in approximately 30 ms on average, which is compatible with the latency requirements of the current real-time transient alert workflow.

\subsubsection{Overall Filtration Performance (Step 1 and 2)}
\begin{figure}[h]
    \centering
    \includegraphics[width=0.6\textwidth]{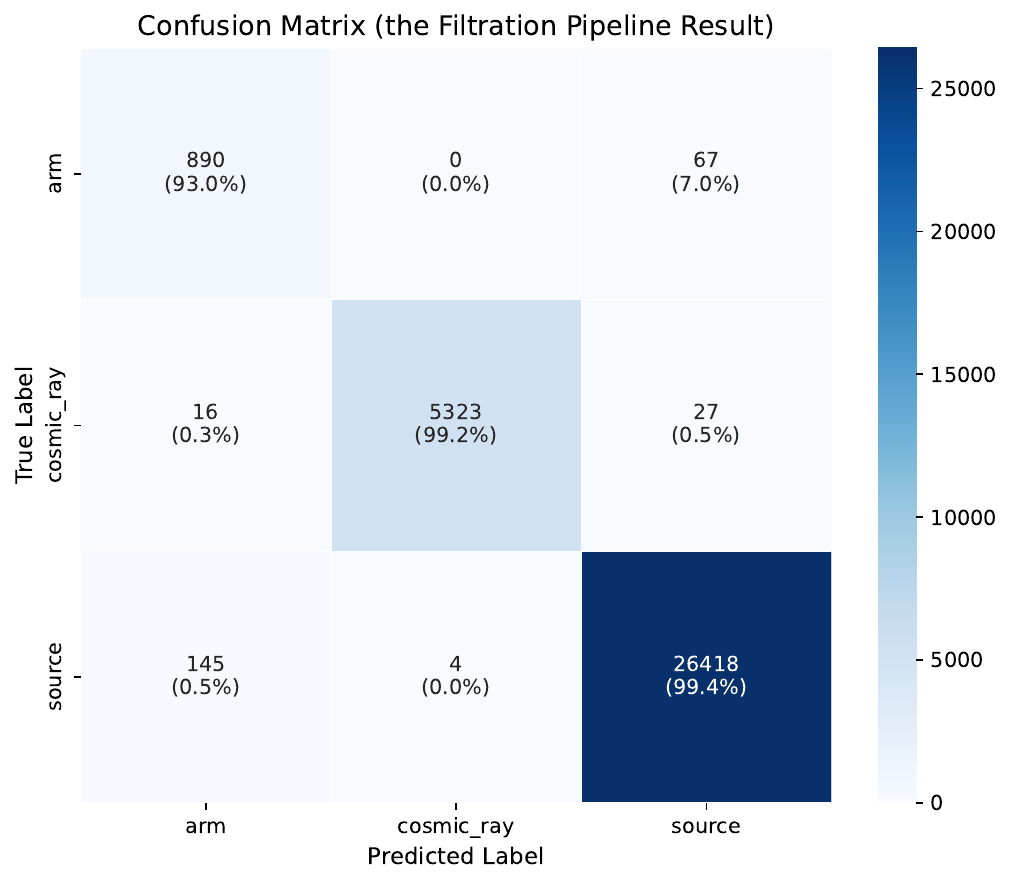} 
    \caption{\textbf{Combined Confusion Matrix of the Filtration Pipeline.} The matrix shows the final classification performance after passing through both Step 1 and Step 2. High diagonal values indicate good separation between real sources and bogus events (Arm/Cosmic Ray).}
    \label{fig:final_cm}
\end{figure}

By cascading the Arm Filter (Step 1) and the Cosmic Ray Filter (Step 2), the pipeline achieves the Real-Bogus classification. 
Figure~\ref{fig:final_cm} illustrates the combined confusion matrix derived from actual intermediate outputs recorded during routine operation of the EP-WXT data pipeline from Dec 5, 2025 to Feb 1, 2026, covering all 32,890 pipeline-generated observations in this period. The cascaded system rejects over 93\% of instrumental Arms and over 99\% of Cosmic Ray events in this operational sample, while maintaining a high recall for genuine sources. This reduction in false positives decreases the manual verification workload required for the scientific operation team.

\subsection{Scientific Validation of Variability Screening (Step 3)}

In pipeline operation, Step 3 functions as the final variability-screening step for observations that survive the preceding filtration steps. Rather than assigning source taxonomy, its practical role is to provide an interpretable basis for rapid manual assessment of statistically significant intra-observation variability.

In practical operation, the Step 3 module generates a Bayesian Blocks diagnostic product for every retained observation. A representative example is shown in Figure~11. The product combines block-wise source, background, and net count-rate information with cumulative source and scaled background counts, providing an interpretable basis for manual vetting and follow-up assessment in archive-scale screening.

\begin{figure}[t]
    \centering
    \includegraphics[width=0.88\textwidth]{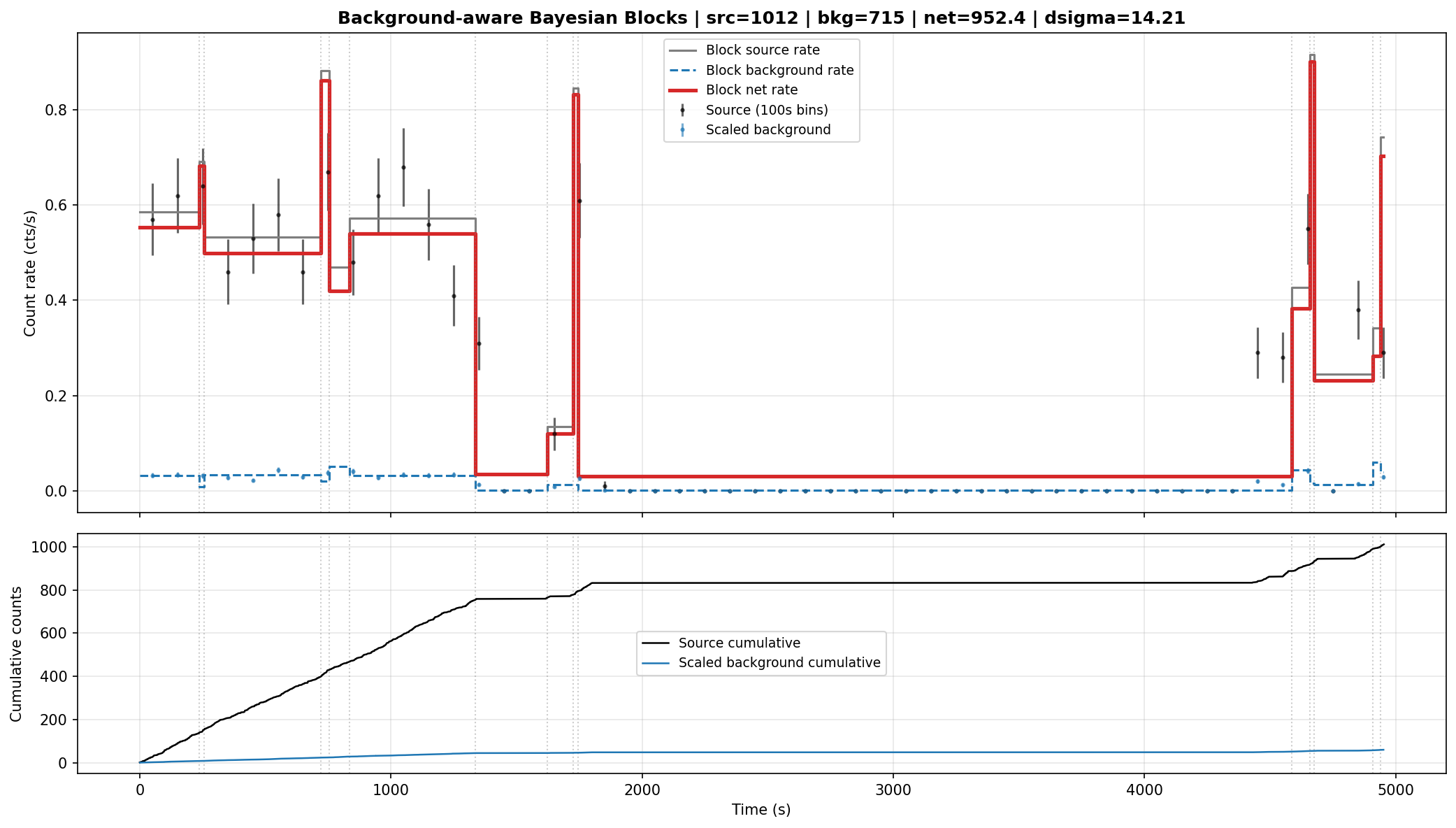}
    \caption{\textbf{Representative Bayesian Blocks diagnostic product for a flagged EP-WXT observation.} The upper panel shows the block-wise source, background, and net count-rate behavior, while the lower panel shows the corresponding cumulative source and scaled background counts. This figure illustrates the standard interpretable output generated by Step 3 for manual inspection of candidate variable observations.}
    \label{fig:bb_example}
\end{figure}

\subsection{Discussion and Future Prospects}
\label{subsec:discussion}

\subsubsection{Impact of Spatial Context and Model Limitations}
A primary source of operational failure in Step 1 is contamination in crowded fields. When a candidate is located near a very bright source, its extended Arm can intrude into the candidate's bounding box, creating complex morphological overlaps that degrade the performance of the Arm filter.

\begin{figure}[t!]
    \centering
    \begin{subfigure}[b]{0.35\textwidth}
        \centering
        \includegraphics[width=\textwidth, height=5.5cm, keepaspectratio]{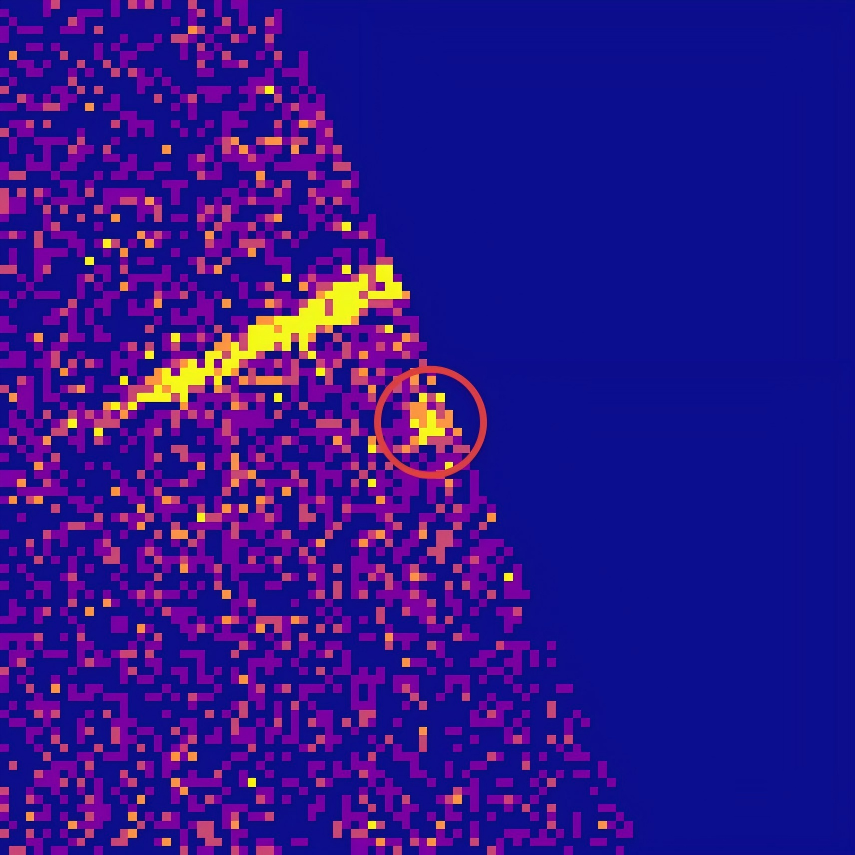}
        \caption{Contamination Case}
    \end{subfigure}%
    \hspace{3em}
    \begin{subfigure}[b]{0.4\textwidth}
        \centering
        \includegraphics[width=\textwidth, height=5.5cm, keepaspectratio]{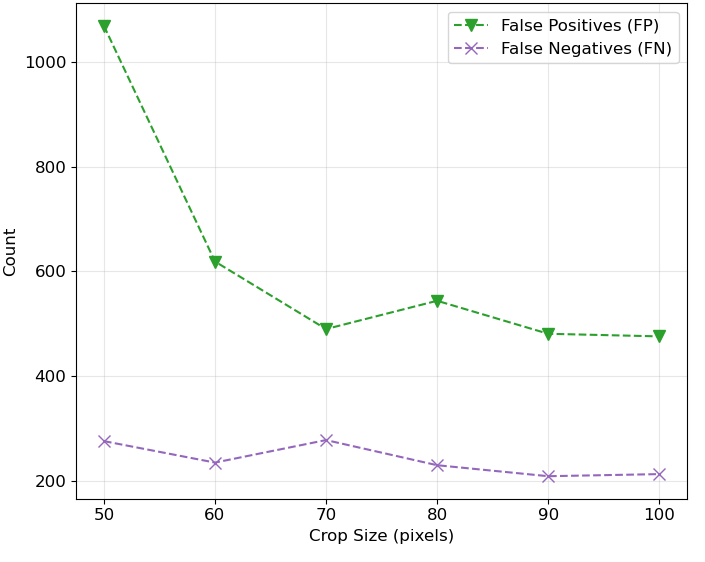}
        \caption{Error Count Trend vs. Crop Size}
    \end{subfigure}
    \caption{\textbf{Analysis of Image Crop Size.} \textbf{(a)} A real source contaminated by the Arm of a neighboring bright source. \textbf{(b)} The trend lines illustrating the sharp decrease in both FP and FN as the crop size increases to $100 \times 100$.}
    \label{fig:crop_size_analysis}
\end{figure}

To investigate the model's sensitivity to the Field-of-View (FoV), we evaluated the ResNet-18 model across crop sizes from $50 \times 50$ to $100 \times 100$ pixels. This ablation study was conducted on a highly imbalanced, independent observation dataset from August to October 2025 (comprising roughly 53,200 real sources and 2,900 Arm artifacts). Because extreme class imbalance can heavily skew percentage-based precision metrics, we present a critical analysis focusing strictly on the absolute error counts: False Positives (FP: Arms misclassified as Sources) and False Negatives (FN: Sources misclassified as Arms). The results are detailed in Table~\ref{tab:crop_size_analysis}. Figure~\ref{fig:crop_size_analysis} visualizes both a representative contamination case and the error-count trend with increasing crop size.

\begin{table}[h]
    \centering
    \caption{\textbf{Error analysis across different image crop sizes on the highly imbalanced dataset.} 
    The $100 \times 100$ crop achieves a Pareto improvement, minimizing both Type I and Type II errors by leveraging global spatial context, though a non-negligible error floor remains.}
    \label{tab:crop_size_analysis}
    \begin{tabular}{ccc}
        \toprule
        \textbf{Crop Size} & \textbf{FP (Bogus $\to$ Source)} & \textbf{FN (Source $\to$ Bogus)} \\
        \midrule
        $50 \times 50$   & 1069 & 276 \\
        $60 \times 60$   & 619  & 235 \\
        $70 \times 70$   & 490  & 278 \\
        $80 \times 80$   & 544  & 230 \\
        $90 \times 90$   & 481  & 209 \\
        $\mathbf{100 \times 100}$& \textbf{476} & \textbf{213} \\
        \bottomrule
    \end{tabular}
\end{table}

These results indicate that crop size affects the absolute error counts. With a restricted FoV of $50 \times 50$, the model loses the global context of the cruciform diffraction pattern and often interprets local Arm fragments as genuine point sources, causing the False Positive count to increase sharply to 1069. As the FoV expands, the network gains more global context on the Arm structure and the spatial relationship between the candidate and surrounding interferers, reducing the False Positive count to 476 and the False Negative count to 213 in the $100 \times 100$ setting used here.

\textbf{Operational Limitations:} While the $100 \times 100$ configuration provides the most favorable trade-off in this analysis, the absolute error counts also highlight an important limitation of the current spatial filtering setup. Passing 476 artifacts introduces a baseline human vetting workload, and vetoing 213 genuine sources (FN) also implies loss of potentially useful scientific information. This veto leakage suggests that single-scale CNNs may approach a performance bottleneck in heavily crowded regions (e.g., the Galactic plane), where multiple overlapping PSFs disrupt morphological symmetry. Future pipeline iterations could address this issue not only by adjusting crop sizes, but also by incorporating the surrounding source catalog to construct masks for known diffraction spikes prior to network inference.

\subsubsection{Towards Fine-grained Classification: The Physical Bottleneck}
Beyond Real-Bogus separation, an additional objective of the pipeline is to classify genuine sources into specific astrophysical categories (e.g., X-ray Binaries, AGN, Cataclysmic Variables). To this end, we extended the Step 2 multi-modal architecture by fusing it with physical parameters derived from power-law spectral fitting—specifically, the column density ($N_H$) and the photon index ($\Gamma$).

Preliminary tests of this augmented model yielded mixed results. Performance on X-ray binaries (XRBs) exceeded 90\% classification accuracy, whereas reliable separation among Active Galactic Nuclei (AGN), Cataclysmic Variables (CVs), and Galaxy Clusters was not achieved. This discrepancy likely reflects the limited information content of the EP-WXT soft X-ray band (0.5--4 keV), within which the spectral shapes and quiescent variability patterns of these softer populations are strongly degenerate. Our current results therefore suggest that fine-grained classification based solely on EP-WXT data is limited, and that future improvements will likely require multi-wavelength priors from external archives such as Gaia and AllWISE.

\section{Conclusion} 
\label{sec:conc}

In this work, we present M-EPDet as a multi-step, multi-modal post-detection framework integrated into the EP-WXT pipeline and tailored to the complex observational characteristics of lobster-eye MPO data. The framework combines a spatial Arm Filter, a temporal-spectral Cosmic Ray Filter, and a Bayesian-Blocks-based variability screening module for candidate prioritization. Its main contributions can be summarized as follows:
\begin{itemize}
    \item \textbf{Spatial filtering:} a ResNet-based Arm Filter that distinguishes incomplete Arm artifacts from genuine cruciform source morphologies.
    \item \textbf{Temporal-spectral filtering:} a dual-branch Cosmic Ray Filter that combines light-curve evolution with PI information to reduce the risk of vetoing real short-timescale events.
    \item \textbf{Statistical screening:} a background-aware Bayesian Blocks module that prioritizes observations with significant intra-observation variability in the low-count regime.
\end{itemize}

Because the pipeline is cascaded and unidirectional, we report decoupled performance metrics for its main steps. M-EPDet achieved a Real-Bogus Recall of 98.31\% ($98.53\% \times 99.78\%$) for genuine astrophysical sources, together with rejection rates of 92.99\% for instrumental Arms and 98.18\% for Cosmic Ray events. In the final step, the background-aware Bayesian Blocks module flagged 2,117 observations out of 282,099 post-filtration observations, corresponding to a retention rate of 0.75\% and a 99.25\% reduction in candidate volume. Deployed as a lightweight CPU-compatible Docker container, the system operates in real time and supports operational candidate vetting in the EP-WXT pipeline.

At the same time, the current framework retains several limitations. Crowded fields remain challenging for the spatial filter, and fine-grained classification of surviving candidates remains limited by spectral degeneracy within the soft X-ray band. Future work will therefore focus on incorporating additional contextual and multi-wavelength information, including catalog-level priors from archives such as Gaia and AllWISE.

The code developed for this study is open source and available at \url{https://github.com/chenlang-china-vo/M-EPDet}.

\begin{acknowledgments}
This work was supported by the Strategic Priority Research Program of
the Chinese Academy of Sciences (XDB0550100), and the National Natural
Science Foundation of China (NSFC; 12403102, 12373110,
12273077). 

This work is based on the data obtained with Einstein Probe, a space mission supported by the Strategic Priority Program on Space Science of Chinese Academy of Sciences, in collaboration with the European Space Agency, the Max-Planck-Institute for extraterrestrial Physics (Germany), and the Centre National d'Études Spatiales (France). 

Computing resources were provided by the
National Astronomical Observatories, Chinese Academy of Sciences. Data
resources are supported by the China National Astronomical Data Center
(NADC), the CAS Astronomical Data Center, and the Chinese Virtual
Observatory (China-VO). This work is also supported by the Astronomical
Big Data Joint Research Center, co-founded by the National
Astronomical Observatories, Chinese Academy of Sciences and Alibaba
Cloud.

During the preparation of this work, the authors used Gemini (Google) to improve the readability and grammatical accuracy of the manuscript. The authors reviewed and edited the output as needed and take full responsibility for the content of the publication. The architectural design, logic, and validation data presented in this paper are entirely the original work of the authors.
\end{acknowledgments}

\begin{contribution}

Lang Chen developed the methodology, conducted the experiments, analyzed the results, and wrote the first draft of the manuscript. 
Yunfei Xu provided primary supervision for the study, guided the methodological design and scientific interpretation, and revised the manuscript. 
Chenzhou Cui contributed to the overall scientific coordination and direction of the project and revised the manuscript. 
Zhen Zhang and Jinhui Xie contributed to code development and technical discussions. 
Yuan Liu, Dongyue Li, and Hui Sun, as collaborators from the EP science team, contributed to scientific discussion, manuscript review, and revision. 
Xiaoxiong Zuo, Shirui Wei and Wujun Shao contributed through scientific discussion and suggestions. 
All authors read and approved the final manuscript.


\end{contribution}

%
\facilities{Einstein Probe (EP)}

\software{
        Astropy \citep{astropy:2013,astropy:2018,astropy:2022},
        NumPy \citep{harris2020array},
        Pandas \citep{reback2020pandas},
        Matplotlib \citep{Hunter:2007},
        PyTorch \citep{paszke2019pytorch},
        scikit-learn \citep{pedregosa2011scikit}
}


\bibliography{ep}
\bibliographystyle{aasjournalv7}



\end{document}